\RequirePackage{fix-cm}
\documentclass[smallextended]{svjour3}       
\smartqed  
\usepackage{graphicx}

\usepackage{soul}
\usepackage{color, colortbl}
\usepackage{tikz}
\usepackage{multirow}
\usepackage[english=american]{csquotes}
\usepackage{subfig}
\usepackage{tcolorbox}
\usepackage{graphics}
\usepackage{hyperref}
\usepackage{url}
\usepackage{amsmath,amssymb,amsfonts}
\usepackage{threeparttable}
\usepackage{adjustbox}

\begin{document}

\title{Bug Characterization in Machine Learning-based Systems\thanks{This work was supported by: Fonds de Recherche du Québec (FRQ), the Canadian Institute for Advanced Research (CIFAR) as well as the DEEL project CRDPJ 537462-18 funded by the National Science and Engineering Research Council of Canada (NSERC) and the Consortium for Research and Innovation in Aerospace in Québec (CRIAQ), together with its industrial partners Thales Canada inc, Bell Textron Canada Limited, CAE inc and Bombardier inc.}
}


\author{Mohammad Mehdi Morovati \and
        Amin Nikanjam \and 
        Florian Tambon \and
        Foutse Khomh \and Zhen Ming (Jack) Jiang
}


\institute{Mohammad Mehdi Morovati
           \and
           Amin Nikanjam
              \and
              Florian Tambon \and
            Foutse Khomh \at
            SWAT Lab., Polytechnique Montréal, Montréal, Canada\\
            \email{\{mehdi.morovati,amin.nikanjam,florian.tambon,foutse.khomh\}@polymtl.ca}\\
           Zhen Ming (Jack) Jiang \at York University, Toronto, Canada \\
           \email{zmjiang@cse.yorku.ca}
}

\date{Received: date / Accepted: date}

\maketitle

\begin{abstract}

Rapid growth of applying Machine Learning (ML) in different domains, especially in safety-critical areas, increases the need for reliable ML components, i.e., a software component operating based on ML. Since corrective maintenance, i.e. identifying and resolving systems bugs, is a key task in the software development process to deliver reliable software components, it is necessary to investigate the usage of ML components, from the software maintenance perspective. Understanding the bugs characteristics and maintenance challenges in ML-based systems can help developers of these systems to identify where to focus maintenance and testing efforts, by giving insights into the most error-prone components, most common bugs, etc. In this paper, we investigate the characteristics of bugs in ML-based software systems and the difference between ML and non-ML bugs from the maintenance viewpoint. We extracted 447,948 \textit{GitHub} repositories that used one of the three most popular ML frameworks, i.e., \textit{TensorFlow}, \textit{Keras}, and \textit{PyTorch}. After multiple filtering steps, we select the top 300 repositories with the highest number of closed issues. We manually investigate the extracted repositories to exclude non-ML-based systems. Our investigation involved a manual inspection of 386 sampled reported issues in the identified ML-based systems to indicate whether they affect ML components or not. Our analysis shows that nearly half of the real issues reported in ML-based systems are ML bugs, indicating that ML components are more error-prone than non-ML components. Next, we thoroughly examined 109 identified ML bugs to identify their root causes, symptoms, and calculate their required fixing time. The results also revealed that ML bugs have significantly different characteristics compared to non-ML bugs, in terms of the complexity of bug-fixing (number of commits, changed files, and changed lines of code). Based on our results, fixing ML bugs are more costly and ML components are more error-prone, compared to non-ML bugs and non-ML components respectively. Hence, paying a significant attention to the reliability of the ML components is crucial in ML-based systems. These results deepen the understanding of ML bugs and we hope that our findings help shed light on opportunities for designing effective tools for testing and debugging ML-based systems.
\end{abstract}

\section{Introduction}
\begin{figure}
    \centering
    \includegraphics[scale=0.33]{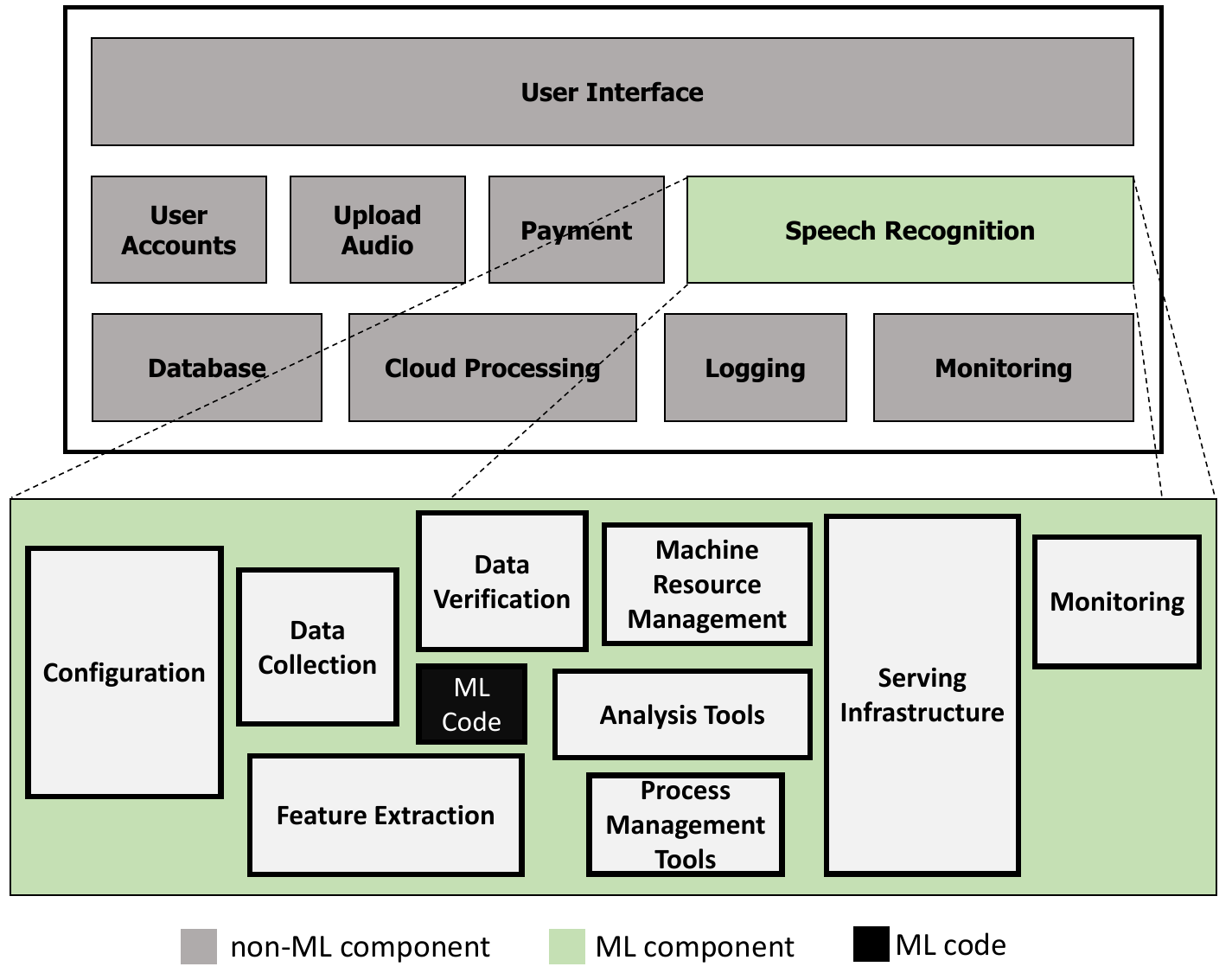}
    \caption{Sample architecture of an ML-based online transcription system (inspired from \cite{sculley2015hidden}).
    }
    \label{fig:ml_based_system}
    \vspace{-1em}
\end{figure}

Machine Learning (ML) is a branch of Artificial Intelligence (AI) that has been applied in a large number of practical applications such as computer Vision~\cite{liu2020small}, and Natural Language Processing (NLP)~\cite{aithal2021automatic}. ML-based systems refer to software systems having at least one ML component~\cite{morovati2023bugs}. A software component is a module or unit of software which is distinct logically or functionally from other modules~\cite{ieee5733835:2010}. Accordingly, an ML component is a software component that operates based on ML. An ML component has several elements affecting its functionality~\cite{morovati2023bugs}.
Figure~\ref{fig:ml_based_system} shows a sample ML-based system; i.e., a transcription service where the ML component provides the key functionality which is converting uploaded voice to text. However, to be able to use the ML model in the production environment, several additional software components, referred to as non-ML components, are necessary. For instance, a user interface is required for enabling users to create a user account, upload the audio file(s), make payment of service expenses, and view the resulting transcript generated from the uploaded audio. Additionally, database and processing components are required to manage related tasks for converting uploaded audio to text, saving transcripts, and a monitoring component to ensure that operational system metrics meet the specified requirements. It is worth mentioning that the ML component itself comprises various elements (from configuration to monitoring) and ML code represents only a small part of the entire ML component~\cite{sculley2015hidden}. 

Software maintenance is the last part of the software development process known for its high required effort and cost~\cite{grubb2003software}. Corrective maintenance, a type of software maintenance focusing on detecting and fixing bugs, is considered as one of the main tasks to deliver reliable software systems~\cite{ieee5733835:2010}. Thus, a clear understanding of software bugs as the core concept of corrective maintenance can assist developers to implement reliable software systems more efficiently. 
There are fundamental differences between traditional software components and ML software components. In traditional software components, developers implement a program that directly encodes the logic to fulfill the functional requirements. However, to develop ML components, one needs to implement an ML model, such as a neural network, and then train it using a dataset~\cite{zhang2020machine}. In other words, ML algorithms learn a complex function from the given dataset and subsequently apply the learned function to the unseen data points. The trained model is then used in the ML components for inference and generating desired results. 
Due to the central role of the trained ML model in the functionality of ML components, we are facing new types of bugs in ML components, referring to the malfunctioning of learned functions for unseen data points, and new challenges in characterizing them~\cite{zhang2018empirical}. 
Accordingly, we can categorize bugs in ML-based systems into two main categories: ML bugs that affect the functionality of ML components and non-ML bugs which affect the functionality of other (non-ML) components. It is important to note that ML models serve as an integral element within ML components that requires interaction with other non-ML and ML components. 

Although a number of studies investigated the bugs in ML programs~\cite{shen2021comprehensive,yan2021exposing}, differences between bug types in ML programs~\cite{humbatova2020taxonomy}, and their symptoms and root causes~\cite{islam2019comprehensive,zhang2018empirical}, none of them provides detailed information on the characteristics of bugs from the software maintenance viewpoint, in comparison with non-ML bugs. This paper aims to fill this gap in the literature by providing insight into the characterization of ML and non-ML bugs in ML-based systems. We also examine the impact of employing ML on the maintenance of software systems. We answer the following research questions:
\begin{itemize}
    \item[\textit{\textbf{RQ1:}}]What are the distributions of the ML and non-ML bugs in ML-based software systems?
    \item[\textit{\textbf{RQ2:}}]What are the differences between the complexity level of fixing ML and non-ML bugs in ML-based systems? 
    \item[\textit{\textbf{RQ3:}}]What is the difference between needed resources (time-to-fix and developer expertise) for fixing ML and non-ML bugs?
\end{itemize}

To answer these questions, we analyzed 386 closed issues gathered from 40 open-source ML-based software systems using the three most popular ML frameworks including \textit{TensorFlow} \cite{abadi2016tensorflow}, \textit{Keras} \cite{chollet2018keras}, and \textit{PyTorch} \cite{paszke2019pytorch}. We summarize the contribution of this study as follows:
\begin{itemize}
    \item We present the first comprehensive empirical study on the characteristics of the bug in ML-based systems, from the software maintenance point of view. 
    \item We provide insight into the cost and complexity of fixing ML and non-ML bugs in ML-based systems.
    \item We indicate and discuss the root causes and symptoms of studied ML bugs.
   \item  We make the dataset used in this study publicly available online~\cite{Replication-Package} for other researchers/practitioners to replicate our results or build on our work.
\end{itemize}

\textbf{The rest of the paper is organized as follows.} We present the background of the study in Section~\ref{sec:background}. Next, we explain the methodology we follow to answer research questions in Section~\ref{sec:methodology}. Section \ref{sec:result} describes the results and findings of the study. 
Then, we discuss the related works and threats to validity of this study in Section~\ref{sec:related_work} and Section~\ref{sec:validity}. Finally, we present the conclusion and future works in Section~\ref{sec:conclusion}.

\section{Background}
\label{sec:background}

\subsection{ML-based systems}

A software component is a self-contained part of software or application that is able to work independently. The reason behind dividing software systems into components is to provide smaller parts with less complexity and more manageability~\cite{lau2018introduction}. The ML community defines ML components as software components that work based on ML algorithms and aim to provide intelligent behavior. ML-based systems are software systems containing at least one ML component~\cite{martinez2021software}.

DL, as a branch of ML, includes neural networks with numerous layers that results in large (deep) models. Recent DL outstanding successes in decision-making and human-competitive tasks (e.g. stock trading~\cite{carta2021multi}) have made it a part of several state-of-the-art software systems ~\cite{humbatova2021deepcrime}. DL-based systems are software systems where the DL model is implemented, trained with a large dataset, and integrated into the software systems~\cite{chen2020comprehensive}. Nowadays, there is an increasing demand for providing reliable DL-based systems, because of the increasing adoption of DL in different software areas, and especially safety-critical areas where any single bug may result in catastrophe (e.g., Tesla autopilot accident~\cite{vlasic2016self}). 

Several frameworks such as \textit{TensorFlow}, \textit{Keras}, \textit{PyTorch}, and \textit{scikit-learn} \cite{scikit-learn} have been provided to help users to design, and implement ML models, and integrate them into the software systems. They provide high-level APIs to simplify the development of ML components for non-expert ML programmers such as software engineers and domain experts~\cite{schoop2021umlaut}. 

\subsection{Bugs in ML-based systems}
Software error is an incorrect part of a source code, which can be a grammatical, logical, or any other type of mistake that a software developer may make \cite{galin2004software}. An ML error is any mistake that occurred in any part of the code of an ML component (e.g. data collection, feature extraction, and ML training code) \cite{zhang2020machine}. A Software system containing erroneous code can produce an abnormal behavior, i.e., a fault (software bug). A software bug is a problem causing a disparity between defined software requirements and developed functionality \cite{ieee5733835:2010}. Accordingly, an ML bug refers to the differences between the existing and required behavior of an ML component \cite{zhang2020machine}.

Based on the location of the bugs (in the code) and their impact, bugs in ML-based systems can be classified into ML and non-ML bugs. We define ML bugs as issues that affect the functionality of ML components. In contrast, a non-ML bug is an issue that deteriorates the functionality of non-ML components. Similar to research conducted by Riccio et al.~\cite{riccio2020testing}, we categorize bugs in ML-based systems into three main classes based on the location of the code where the bugs occur (Figure~\ref{fig:ml_bug}):

\begin{figure}
    \centering
    \includegraphics[scale=0.5]{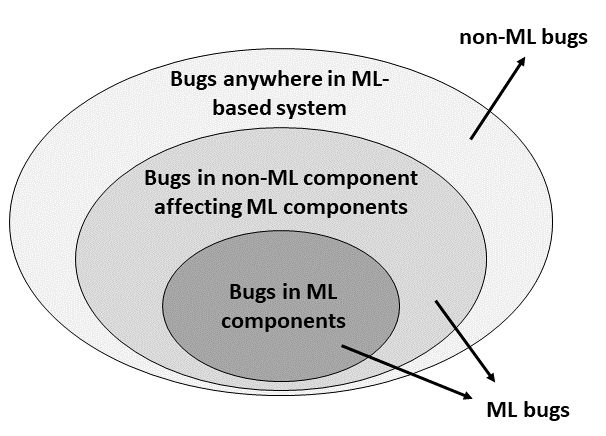}
    \caption{
    Venn diagram of bug categorization in ML-based systems based on the location they occurred.
    }
    \label{fig:ml_bug}
    \vspace{-1em}
\end{figure}

\begin{itemize}
    \item \textit{\textbf{Bugs in ML components}}: In this class, we investigate only bugs inside ML components in isolation. As an example, issue~\cite{issue_sample_01} is related to a problem in training logs, hence it is related to an issue inside the ML component. 
    \item \textit{\textbf{Bugs in non-ML components, but affecting ML components}}: This class considers bugs inside any component affecting ML components' functionality. This category includes all bugs classified into the previous category (Bugs in ML components) as well. For example, issue~\cite{issue_sample_02} explains a problem in the output server channel causing a bug in interactive learning. Although this issue is not inside the ML component, it results in failure in the inference of the ML component.
    \item \textit{\textbf{Bugs anywhere in the system}}: This class contains any bugs in all the system's components, related to ML components or not. An issue of the Rasa project (\#4142)~\cite{issue_sample_03} is an example of a non-ML bug referring to a problem in the presentation of the output.
\end{itemize}

\begin{figure}
    \centering
    \includegraphics[width=0.65\columnwidth]{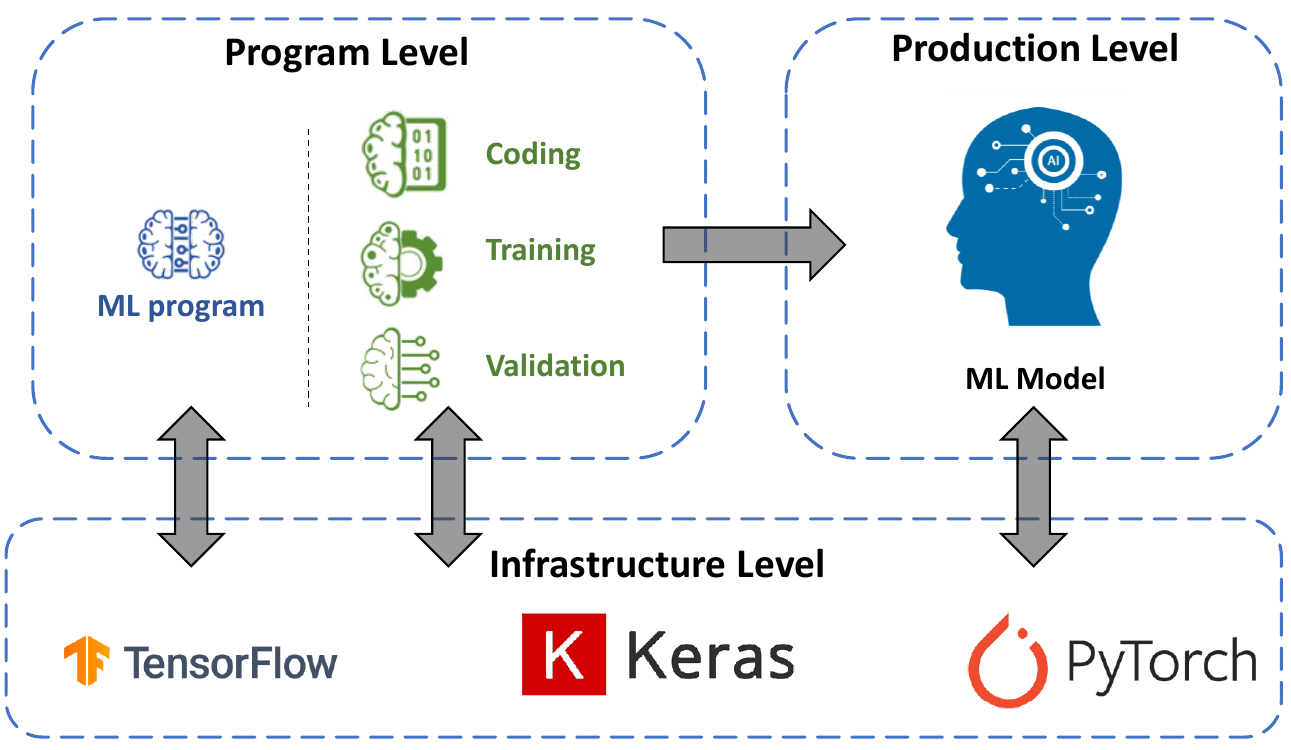}
    \caption{High-level architecture of ML components. 
    }
    \label{fig:bug_dl}
    \vspace{-1em}
\end{figure}

ML bugs can emerge from three main sources: program level (ML program written to build ML model), production level (i.e., using the ML model for inference), and infrastructure level (ML frameworks) (see Figure \ref{fig:bug_dl}). Bugs at any level may also affect the overall quality of the ML component and accordingly, the quality of the ML-based system. In this study, we focus on the ML bugs at the program level. 


When a user tries to use a buggy section of the software, it triggers the bug which results in a system failure. System failure refers to the inability of the software system to perform its identified functionality. In addition to the general concept of system failure, ML bugs may also result in a bad performance, crash, data corruption, hang, and memory out of band which are considered ML failures as well~\cite{islam2019comprehensive}. 

\subsection{SRE in ML-based systems}
SRE is the study of the functional behavior of the software systems in conformance with the user requirements, in terms of reliability~\cite{lyu2007software,IEEE:reliability:7827907}. Software reliability implies the ability of the software system or component to perform required functionality during a specific period of time. Essential differences between the paradigms of traditional and ML-based software systems generate new challenges in SRE of ML-based systems. Fast changes in new versions of ML frameworks~\cite{islam2020repairing}, unportable code~\cite{lenarduzzi2021software}, bug reproducibility~\cite{zhang2018empirical}, and lack of detailed information about bugs~\cite{wardat2021deeplocalize} are the most significant challenges in SRE of ML-based systems. Thus, SRE methods working effectively for traditional software systems operate unsatisfactory for ML-based systems. As such, SRE techniques need to be adapted from the ones originally developed for traditional software systems to ML-based systems.  

One of the most significant SRE tasks to enhance the reliability of the software system and stop the recurrence of software failures is software maintenance~\cite{wang2006reliability}. Software maintenance is the process of modifying software systems after delivering them to the end user, in order to fix discovered bugs, improve software performance, or add new features to adapt the software to new requirements. Software maintenance is classified into four basic types including adaptive, corrective, perfective, and preventive~\cite{ieee5733835:2010}. Corrective maintenance is an essential part of SRE that aims at fixing bugs after delivering the software systems to the users~\cite{IEEE:reliability:7827907}. 
Concerning that characterizing bugs is the first step toward bug fixing~\cite{ni2020analyzing}, bug characterization is considered a necessary stage in the maintenance process.
Thus, we are going to shed light on the way that ML and non-ML bugs affect the corrective maintenance of ML-based software systems.
Besides, understanding bug characterizations potentially improves developers' expertise in debugging and development practices. In other words, our results would help ML-based software developers to identify best practices that lead to fewer bugs. Moreover, the provided results could foster the implementation of automated testing tools for bug detection, bug localization, and debugging.

\section{Methodology}
\label{sec:methodology}

\begin{figure*}
    \centering
    \includegraphics[width=\textwidth]{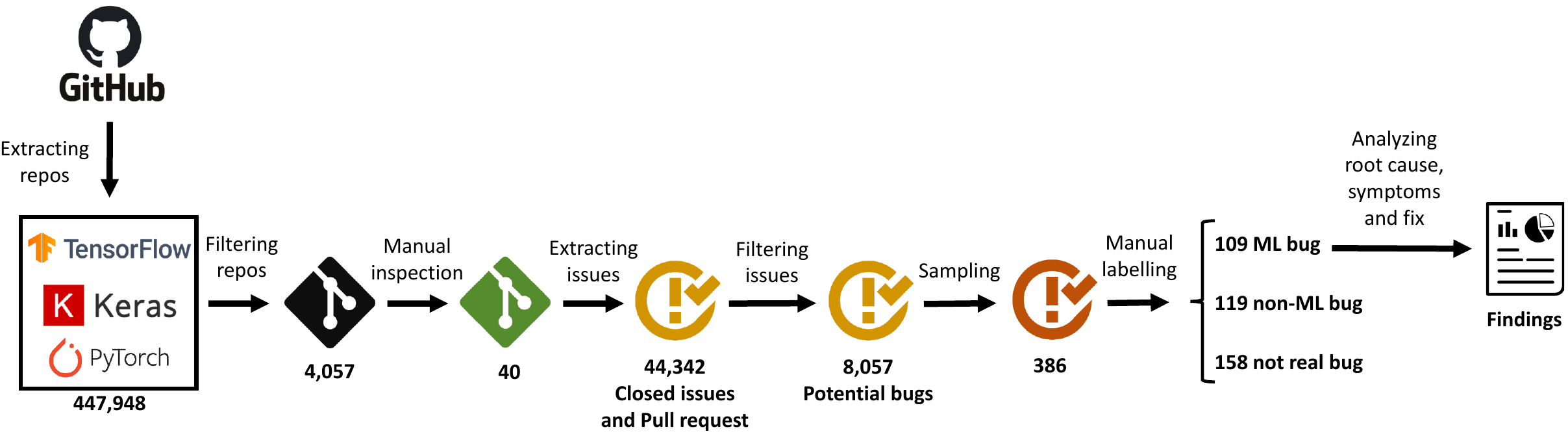}
    \caption{High-level view of the followed methodology.} 
    \label{fig:method}
\end{figure*}

In this section, we describe our methodology to investigate bugs in ML-based software systems answering our RQs. The general process consists of three steps: (1) Collecting repositories, (2) Manual inspection of repositories, and (3) Manual labeling of bugs. Figure~\ref{fig:method} illustrates the methodology that we followed in our study. 

\subsection{Collecting repositories}
To identify relevant ML issues, this study focuses on projects developed using the three most popular ML frameworks: \textit{TensorFlow}, \textit{Keras}, and \textit{PyTorch}. We select them based on their popularity metrics in \textit{GitHub} as shown in Table~\ref{tbl:ml_frameworks}, e.g., the number of stars and forks.

\begin{table}
\small
\caption{Details of the selected ML frameworks.}
    \centering
    \begin{tabular}{p{3.5cm} r r r}
        \hline
            \textbf{ML Framework} & \textbf{\#stars} & \textbf{\#forks} & \textbf{\#subscribers} \\
            \hline
            \rowcolor{gray!40}
            TensorFlow & $174 k$ & $88.3 k$ & $3.4 k$\\
            \rowcolor{gray!40}
            Keras & $58.3 k$ & $19.3 k$ & $2 k$\\
            \rowcolor{gray!40}
            PyTorch & $66.7 k$ & $18.3 k$ & $1.7 k$ \\
            Caffe & $ 33.3 k$ & $19 k$ & $269$ \\
            Jax & $ 23.1 k$ & $2.2 k$ & $504$ \\
            MXNet & $ 20.4 k$ & $6.9 k$ & $875$ \\
            CNTK & $ 17.4 k$ & $4.4 k$ & $201$ \\
            sonnet & $ 9.6 k$ & $1.4 k$ & $51$ \\
        \hline
    \end{tabular}
        \label{tbl:ml_frameworks}
        \vspace{-1em}
\end{table}

We use \textit{GitHub} as the main source to extract the needed data for our analysis. As of December 2021, \textit{GitHub} \cite{github-website} has more than 73 million registered users and over 200 million repositories. Computer society also considers \textit{GitHub} as the most important resource of open source software systems~\cite{li2020exploratory}. \textit{GitHub} provides APIs to facilitate extracting detailed data about repositories. We use \textit{GitHub} search API~\cite{github_api_v3} to extract repositories using identified ML frameworks. \textit{GitHub} search API receives a query as input and returns a list of repositories that match the query.

Since \textit{Python} is the most popular programming language for ML~\cite{voskoglou2017best,Gupta:MLLangugae}, and \textit{TensorFlow}, \textit{PyTorch}, and \textit{Keras} provide \textit{Python} APIs, we also narrow down the search to the repositories developed using \textit{Python} by adding `\texttt{language:python}' parameter to the search queries. In \textit{Python}, when a developer wants to use a library, she adds the library to the script using the `\texttt{import}' command before its usage. For example, developers add `\texttt{import keras}' to the script to use \textit{Keras} in their application. So, we run search queries with `\texttt{import <ML framework>}' (\textit{TensorFlow}, \textit{Keras}, and \textit{Pytorch}) to fetch all repositories using these frameworks. Since \textit{Keras} may be imported using `\texttt{import tensorflow.keras}', we also search for this command as well.
\textit{GitHub} limits the users to access only the first 1000 results of the search API. To overcome this issue, we manage to perform different search queries by adding a parameter to fetch files with the size inside the indicated range (i.e., \texttt{size:<min>..<max>}). We define the range for each query in such a way that achieves less than 1000 results. Because our query returns an empty list for the repositories using the files greater than 500 \textit{MB}, we divide the whole range between 1 \textit{KB} and 500 \textit{MB} into ranges of 10 \textit{KB}. So, we raise 150,000 search API calls in total, 50,000 for each framework to fetch repositories. 

\begin{table}
\small
\caption{Detailed information about the number of remaining repositories after each filtering out step.}
    \centering
    \begin{tabular}{p{6cm} r r r}
        \hline
            \multirow{2}{6em}{} & \multicolumn{3}{ c }{ML frameworks} \\
             \cline{2-4}
            & \multicolumn{1}{c}{\textbf{\textit{TensorFlow}}} & \multicolumn{1}{c}{\textbf{\textit{Keras}}} & \multicolumn{1}{c}{\textbf{\textit{PyTorch}}} \\ 
            \hline
            All extracted & 144415  & 98462 & 205071 \\
            After removing unpopular & 10480  & 4108  & 12556\\
            After removing inactive & 4393 & 2989 & 7171\\
            After removing repos with trivial history & 1633  & 738 & 2189 \\
            After removing personal & 1445 & 661  & 1950 \\
            Repos with ML keywords & 156 & 488 & 250\\
            After manual checking top 100 & 12 &  15 & 13\\
            
            \hline
            \hline
            \textbf{Total extracted repos} & \multicolumn{3}{r}{\textbf{447948}} \\
            \textbf{Total remained repos} & \multicolumn{3}{r}{\textbf{4057}} \\
            \hline
            \textbf{Identified ML-based repos} & \multicolumn{3}{r}{\textbf{40}} \\
            \hline
    \end{tabular}
        \label{tbl:repo_filter}
        \vspace{-1em}
\end{table}

Next, we conduct four filtering steps based on repositories' metadata to remove personal, inactive, and unpopular repositories, and ones with trivial history (Table~\ref{tbl:repo_filter}). For mining repositories' metadata such as the number of open/closed issues, and the number of commits, we use \textit{GitHub} GraphQL API~\cite{github_GraphQL_API} that outperforms API V3 for extracting such information. In other words, we replace several \textit{GitHub} API V3 calls with one GraphQL API call. We use the following exclusion criteria that were employed successfully in previous studies~\cite{humbatova2020taxonomy,krishna2018connection,hata2021same}: 
\begin{itemize}
    \item \textbf{unpopular repositories}: repositories with less than 10 stars or 10 forks. 
    \item \textbf{personal repositories}: repositories with 1 collaborator.
    \item \textbf{inactive repositories}: repositories without any activity during last year. 
    \item \textbf{repositories with trivial history}: repositories with less than 100 commits.
\end{itemize}

We are left with 4,057 repositories after this step.

\subsection{Manual inspection of repositories}
After collecting repositories and filtering out irrelevant ones, we proceed with manual inspection of them. This is necessary to identify true ML-based systems. In the first step, we selected 30 repositories randomly (10 for each framework) with the highest number of closed issues and PRs similar to the methodology followed by Humbatova et al.~\cite{humbatova2020taxonomy}. We then manually check all 30 repositories to ensure that they are ML-based systems and have at least one ML component. We establish the following exclusion criteria to investigate collected repositories and remove non-ML-based repositories:
\begin{itemize}
    \item Repositories that use a language other than English to describe their application, such as~\cite{github_repo_chinese},
    \item Tutorial and training repositories which are a set of sample codes or implemented examples of books, such as~\cite{github_repo_training},
    \item Repositories using ML framework, but not for implementing the functionality of components based on ML algorithms (Figure \ref{fig:bug_dl}), such as \cite{github_repo_data_preprocess} that uses \textit{Keras} for data preprocessing,
    \item Repositories using ML frameworks for testing their functionality or providing examples of their application usage, such as~\cite{github_repo_test_func},
    \item Repositories that implement a wrapper for the ML frameworks to extend their functionality, such as~\cite{github_repo_wrapper}.
\end{itemize}
To proceed, the first three authors went through 30 randomly selected repositories and checked them separately, to categorize them into ML-based and non-ML-based, using an open coding procedure~\cite{seaman1999qualitative}. After a meeting to discuss the results, they concluded that only 3 out of the 30 manually checked repositories are ML-based systems. Hence, we added another filtering criterion to decrease the number of false positives. We investigate repositories’ scripts for keywords related to defining, training, and evaluating ML models and exclude repositories without any of such keywords. For instance, in \textit{Keras}, \textit{`Sequential'} and \textit{`Model'} APIs are used to define the model, and \textit{`fit'}, \textit{`compile'}, \textit{`evaluate'}, \textit{`predict'}, \textit{`train\_on\_batch'}, \textit{`test\_on\_batch'}, \textit{`predict\_on\_batch'}, and \textit{`run\_eagerly'} APIs for training and evaluating models~\cite{keras_doc}. We followed the same approach for \textit{TensorFlow} and \textit{PyTorch}, and reported details in our replication package~\cite{Replication-Package}.
By running this filtering step, we retained 894 repositories.

We then selected a sample of 100 repositories for each framework, in total 300. The top 100 repositories with the highest number of closed issues and PRs have been selected for each ML framework, similar to the methodology followed by Humbatova et al.~\cite{humbatova2020taxonomy} to be checked manually. To label repositories into ML-based or non-ML-based, the first three authors classified the first 15 repositories of each ML framework (45 in total) independently. To assess inter-rater agreement among them, we used Fleiss’ kappa~\cite{falotico2015fleiss} like in similar works~\cite{yang2022mining,quach2021empirical}, and obtained an inter-rater agreement of about 33\%. Next, the three authors meet to discuss the conflicts and resolve them. Afterward, we repeated the labeling process for 45 repositories and the agreement rate reached 89.9\%, which is acceptable to keep going through the rest of the repositories (Fleiss’ kappa agreement greater than 81\% is interpreted as almost perfect agreement~\cite{hartling2012validity}). So, we proceeded to label the rest of the repositories, and the three authors labeled all 300 repositories and achieved 87.7\% agreement. In the end, 40 repositories were identified as ML-based. The list of repositories is available in our replication package~\cite{Replication-Package}. The list encompasses projects such as AirSim\footnote{\url{https://github.com/microsoft/AirSim}} a simulator for drones and cars, FaceSwap\footnote{\url{https://github.com/deepfakes/faceswap}} a deep learning tool to swap faces on videos or DeepSpeech\footnote{\url{https://github.com/mozilla/DeepSpeech}} a speech-to-text engine.

It is worth noting that we do not consider any exclusion criterion to filter out any ML algorithm. However, after manual inspection of repositories, all of the remaining repositories are DL-based software systems. Besides, with respect to the fact that we aim at comparing characteristics of bugs in ML-based vs non-ML-based systems in this study, we do not make any distinctions between various types of ML algorithms (CNN, Transformer, RNN, etc.).

\subsection{Manual labeling of bugs}
Since our goal in this study is to characterize bugs in ML-based systems, following the existing work \cite{shen2021comprehensive,nikanjam2022faults}, we mine the 40 identified ML-based repositories and extract closed issues and merged Pull Requests (PR) showing a bug-fix. We choose such PRs because 1) bugs mentioned in such PRs have been already accepted and then got fixed and 2) these PRs usually have more comprehensive information about the fixed bugs (e.g., code changes, links to related issues, and discussions among developers), which facilitates understanding the bugs. To identify PRs with the purpose of bug-fixing from repositories, following the existing study \cite{shen2021comprehensive,garcia2020comprehensive}, we collect PRs whose tags/titles include at least one bug-relevant keywords (i.e., fix, defect, error, bug, issue, mistake, incorrect, fault, and flaw). Similarly for closed issues, as users raise \textit{GitHub} issues for several purposes (e.g. asking questions, enhancement, feature request, and reporting bugs) and assign some tags to represent their goal, closed issues with at least one of the mentioned keywords in tags or titles were extracted.

We initially extracted a total of \textbf{44,342} closed issues and merged PRs. Out of them, we achieved \textbf{8,057} with the bug-relevant keywords. Following previous works \cite{chen2020comprehensive,zhang2019}, to ensure a 95\% confidence level and a 5\% confidence interval, we randomly sample \textbf{367} issues/PRs. With respect to the different number of raised issues/PRs in each ML-based repository of our dataset, we use the Stochastic Universal Sampling (SUS)~\cite{wirsansky2020hands} method to obtain a fair set of sampled artifacts and prevent any bias against the repositories with the small number of issues/PRs. In other words, we extract bugs from each repository, based on the ratio of each repository’s issues/PRs to the total number of extracted ones. We round calculated values to the closest greater integer number, and that left us with \textbf{386} issues/PRs at the end.

In the next step, we inspect the extracted issues/PRs manually to identify their type as \textit{ML} or \textit{non-ML}, their root causes and symptoms following an open coding procedure \cite{seaman1999qualitative}. To this end, we use a two steps manual labeling: 1) to categorize the issues/PRs into ML and non-ML and, 2) to classify the ML ones based on their root causes and symptoms. To categorize the issues/PRs into ML and non-ML, each of the first three authors went independently through the first 40 issues/PRs (almost 10\% of all sampled issues) as a pilot analysis, and we achieved a 44.8\% inter-rater agreement Fleiss' kappa. To identify the main reasons for disagreements and resolve them, two meetings were held and a clear criterion was agreed upon for each class. Generally, we label issues/PRs affecting the quality of ML components as ML bugs and the rest as non-ML bugs. During the manual labeling, several collected closed issues turned out not to be real issues. As an example, some issues are generated automatically by the repository's bot~\cite{github_issue_bot_generate} or closed by it, because nobody replied to the issue~\cite{github_issue_bot_close}. So, we added an additional group as '\textit{not real bug}' referring to the issues which can be classified as neither ML nor non-ML, satisfying one of the following rules:

\begin{itemize}
    \item issues generated automatically by a bot.
    \item issues which are end-user questions (users do not know how to handle the problems) or user's mistakes.
    \item issues that are closed without fixing (not having enough information from the issue report or having a rejected PR).
    \item issues that are not reproducible.
    \item issues that are unclear or without explanation.
\end{itemize}

We also filter out the PRs that are found to be irrelevant to bug fixing (labeled as \textit{no real issue}). Moreover, some PRs fixed multiple bugs, and we, therefore, labeled each of them as an individual bug where applicable, similar to the existing work \cite{garcia2020comprehensive,shen2021comprehensive}.
 
Then we labeled those 40 issues/PRs again and achieved an 89.6\% agreement (interpreted as almost perfect agreement~\cite{hartling2012validity}), which is reasonable to keep labeling the rest of the issues/PRs. To label the remaining issues/PRs, we label every 100 bugs, have a meeting to discuss the results, explore the main reasons for disagreements, and resolve them for the next parts. Eventually, we achieve \textbf{89.7\%} agreement, resulting in \textbf{109} ML bugs (28.2\%), \textbf{119} non-ML bugs (30.8\%) and \textbf{158} not real bugs (40.9\%). Table~\ref{tbl:issues_detailed} illustrates the detailed information regarding manual labeling of issues as ML bugs or non-ML bugs.  

\begin{table}
\footnotesize
\caption{Detailed information about labeled bugs.}
    \centering
    \begin{tabular}{p{2cm} r r | r r r}
        \hline
            \textbf{Framework} & \multicolumn{1}{c}{\textbf{\textit{Issue}}} & 
            \textbf{\textit{PR}} &
            \textbf{\textit{ML bug}} &
            \multicolumn{1}{c}{\textbf{\textit{non-ML bug}}} & 
            \multicolumn{1}{c}{\textbf{\textit{Not real bug}}}\\
            \hline
            TensorFlow & 88 & 32 & 18 & 41 & 61 \\
            Keras & 76 & 11  & 8 & 46 & 33\\
            PyTorch & 118  & 61 & 83  & 32 & 64\\
            \hline
            \textbf{Total} & \textbf{282}  & \textbf{104} & \textbf{109} & \textbf{119} & \textbf{158}\\
        \hline
    \end{tabular}
        \label{tbl:issues_detailed}
        \vspace{-2em}
\end{table}

Then, we carry out the second part of bug labeling to indicate their \textbf{root cause} and \textbf{symptoms}. Since many of our selected repositories used DL models, we employ the recent classification of root causes and symptoms of bugs in DL systems introduced in \cite{islam2019comprehensive} which extended a former study \cite{zhang2018empirical}. For root causes, we have: \textit{absence of inter API compatibility} (AIAPIC), \textit{absence of type checking} (AOTC), \textit{API change} (APIC), \textit{API misuse} (APIM), \textit{confusion with computation model} (CWCM), \textit{incorrect model parameter or structure} (IMPS), \textit{structure inefficiency} (SI), \textit{unaligned tensor} (UT), and \textit{wrong documentation} (WD). 
Table~\ref{tab:bug_root_cause} represents the detailed description of root causes of ML bugs in ML-based systems.
The types of symptoms are as follows: \textit{bad performance}, \textit{crash}, \textit{data corruption}, \textit{hang}, \textit{incorrect functionality}, and \textit{memory out of bound}. We mainly rely on the definitions presented in their original sources \cite{zhang2018empirical,islam2019comprehensive}.
To label the bugs, we first selected 20 ML bugs out of the 109 ML bugs and labeled them as a pilot step, achieving 30\% and 84.9\% agreements based on Fleiss' kappa for root causes and symptoms, respectively. We again meet two times to identify the major reasons for disagreements and reach a consensus on the labels' concepts and definitions. Then, labeling the first 20 ML bugs again resulted in an 86.5\% inter-rater agreement for root causes and 89.7\% for symptoms. For the rest of the bugs, we inspected them in three rounds. In each round, we label around 1/3 of the bugs, then meet to discuss the disagreements and resolve them. In the end, we achieved \textbf{88.4\%} and \textbf{97.2\%} agreement for root causes and symptoms of ML bugs.

\begin{table}[]
    \centering
    \resizebox{\columnwidth}{!}{
    \begin{tabular}{p{3cm} p{10cm}}
    \hline
     \rowcolor{gray}
     \textbf{Bug root cause}& \textbf{Description} \\
     \hline
     Absence of Inter API Compatibility (AIAPIC) & Inconsistency of two different types of libraries. For example, issue!\cite{issue_sample_08} emerges from a problem in using \textit{`pytest'} in the application developed using PyTorch.\\
     \rowcolor{lightgray}
     Absence of Type Checking (AOTC) & Data type mismatch in calling API methods. As an example, issue~\cite{issue_sample_09} stems from the absence of type checking of “trainer.logger” in different places. \\
     API Change (APIC) & Releasing new versions of ML frameworks that their APIs are not compatible with their previous versions. Issue~\cite{issue_sample_10} that is raised because of API changes in versions 1.2 and the older ones of pytorch lightning is considered as an example of APIC. \\
     \rowcolor{lightgray}
     API Misuse (APIM) & Trying to use an API developed by an ML framework, without a sound understanding about it. \\
     Confusion with Computation Model (CWCM) & Confusion about the function of an API of the ML framework leading to the misuse of the assumed computation model by the ML framework. \\
     \rowcolor{lightgray}
     Incorrect Model Parameter or Structure (IMPS) & Constructing the ML model such as creating an ML model with  incorrect structure or using inappropriate parameters. For instance, issue~\cite{issue_sample_11} raises because of ignoring parameters’ default value of the early stopping method. \\
     Structure Inefficiency (SI) & Problems in the modeling step of ML component/software. Although this category is similar to IMPS, their symptoms are different. In fact, IMPS leads to system crashes, however SI results in system bad performance. Issue~\cite{issue_sample_12} is an example that its root cause is using incorrect training parameters resulting in system bad performance. \\
     \rowcolor{lightgray}
     Unaligned Tensor (UT) & Trying to build a computation graph in an ML process without providing input data satisfying input specification of the ML framework API. For instance, issue~\cite{issue_sample_13} stems from the shape mismatch of the intermediate tensors. \\
     Wrong Documentation (WD) & Incorrect information presented in the formal documentation of the library. As an example, issue~\cite{issue_sample_14} is related to the unclear explanation in the document of the lithening\-AI library for using early stopping API. \\
     \hline
    \end{tabular}
    }
    \caption{
    The identified root causes of ML bugs in open-source ML-based software systems.
    }
    \label{tab:bug_root_cause}
\end{table}

\section{Empirical Results}
\label{sec:result}
In this section, we provide and discuss empirical results for each RQ. The materials and all collected data used for this study are publicly available in our replication package \cite{Replication-Package}.

\subsection{RQ1: Distribution of bugs in ML-based Systems}\label{sec:rq1}

\begin{figure}
  \centering
  \subfloat[]{\includegraphics[width=0.29\textwidth]{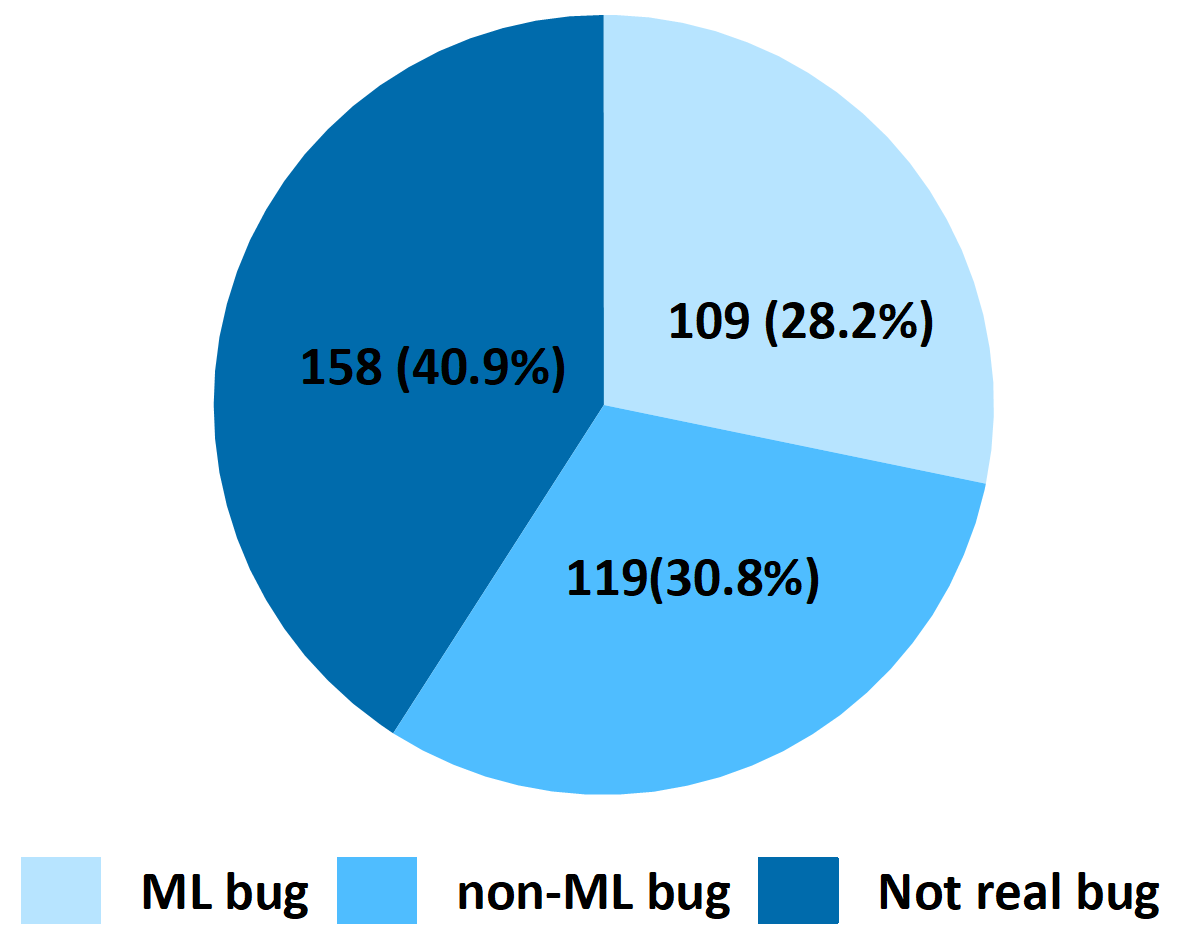}\label{fig:bug_distribution}}
  \subfloat[]{\includegraphics[width=0.29\textwidth]{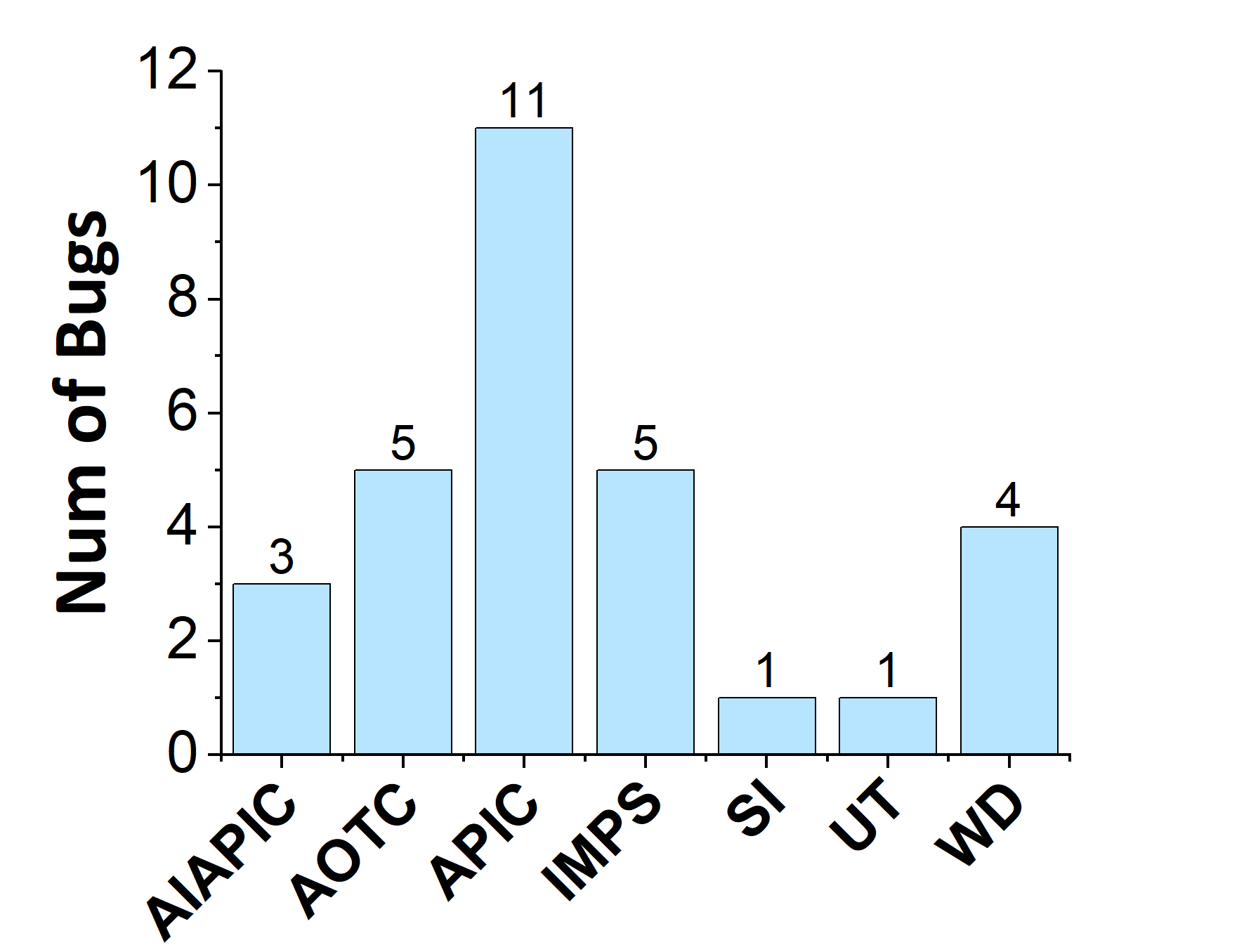}\label{fig:bug_rootCause}}
  \subfloat[]{\includegraphics[width=0.31\textwidth]{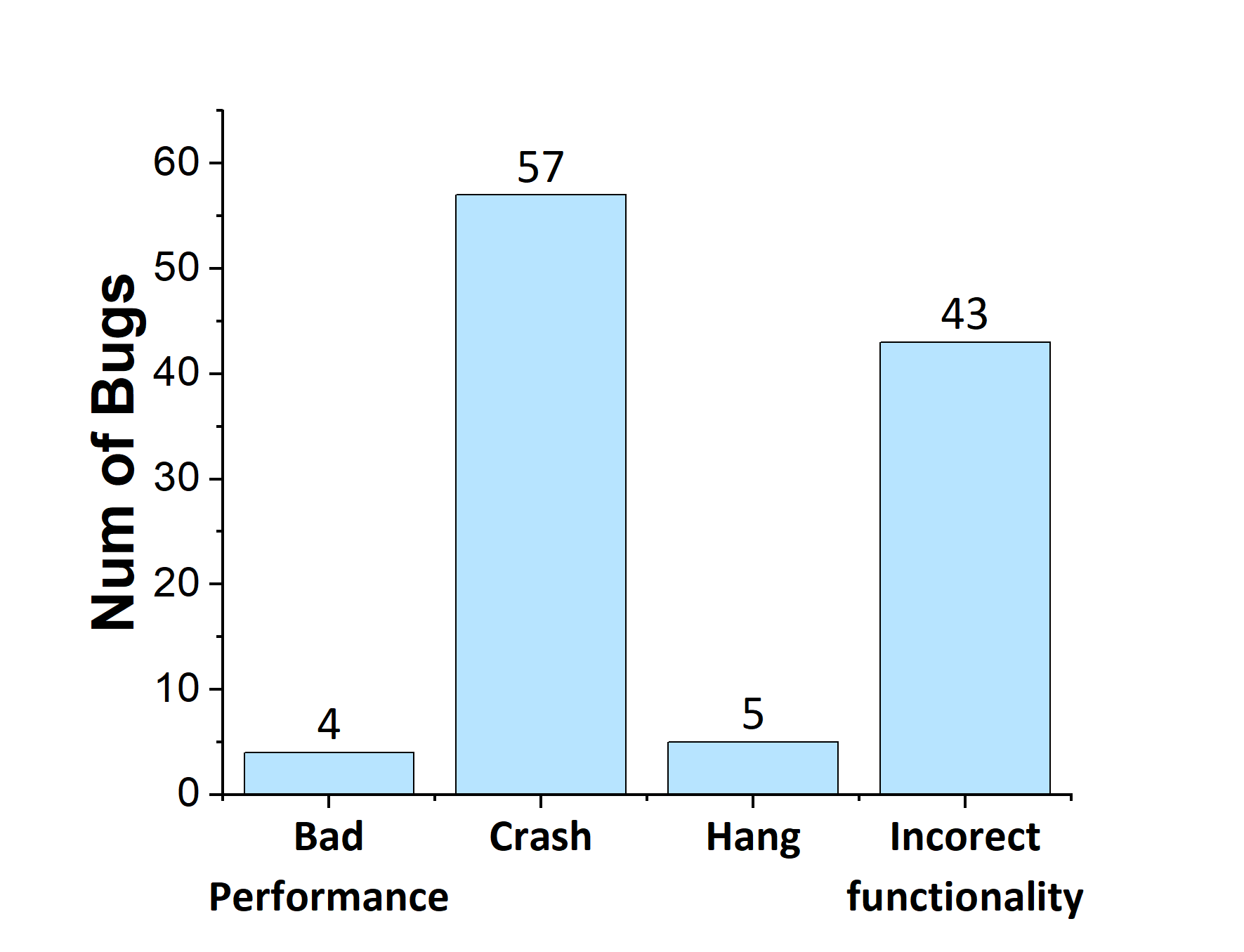}\label{fig:bug_symptoms}}
  \caption{Bug in ML-based systems: (a) distribution of issues, (b) root causes of the ML bugs, and (c) symptoms of ML bugs. 
}
\label{fig:dist}
 \vspace{-1em}
\end{figure}

Our goal in RQ1 is to evaluate the distribution of ML and non-ML bugs in ML-based software systems. Results are presented in Figure~\ref{fig:dist}. Manual labeling of a selected sample (with a 95\% confidence interval) of potential bugs reveals that 40.9\% of them are not real bugs. This finding suggests that almost half of the issues/PRs reported in the collected repositories are not real bugs. Some of these issues are identified as users' questions/mistakes while others lack sufficient description to be considered as a bug (e.g., not reproducible or have unclear description/not enough information). This finding aligns with a previous research conducted by Anvik et al.~\cite{10.1145/1117696.1117704} which found that only 58\% of the issues raised for Eclipse and 44\% of the issues raised for Firefox have been either fixed or considered for fixing. On the other hand, Long et al.~\cite{Long22} carried out research on the performance and accuracy bugs within ML frameworks (e.g. TensorFlow, Keras, PyTorch, etc) and reported only 3\% of the raised issues as bugs without sufficient information or non-reproducible. This discrepancy can be explained by the fact that our study encompasses a broader range of bug types in ML-based systems, beyond just accuracy and performance issues, which seem to be more easily identifiable (and likely fixed) than other types of bugs. Furthermore, we focus on the bugs within ML-based systems rather than inside the ML framework itself. Moreover, it is worth noting that localizing and identifying the root causes of bugs inside ML-based systems are likely more challenging than bugs inside ML frameworks, mainly due to the utilization of various libraries in the implementation of ML-based systems. Among the remaining bugs, our results demonstrate that 47.8\% of the real bugs in ML-based systems are dealing with ML components (Figure \ref{fig:bug_distribution}). Although ML components of software take just about 10\% of development time~\cite{menzies2019five}, it represents roughly half of the real issues encountered in a typical ML-based system. These findings emphasize the critical need for automatic testing tools for ML components and ML-based software systems.

\begin{tcolorbox}
\textbf{Finding 1. }A large number of issues in ML-based systems either turn out to be users' questions/mistakes or did not provide enough information to decide whether it is a bug or to help fix it. Moreover, although almost 10\% of development time is consumed for ML components, nearly half of the "real" bugs in ML-based systems are related to ML components. 
\end{tcolorbox}

As mentioned in Section \ref{sec:methodology}, to classify the root cause and symptom of our bugs, we employed the categories introduced by Islam et al. ~\cite{islam2019comprehensive}. Figure \ref{fig:bug_rootCause} presents the indicated root causes. The most important observation is that 72.5\% of the identified ML bugs could not be classified in any of the 9 root causes defined in \cite{islam2019comprehensive}, so we classified them as \textit{Other} (we do not report this category in Figure \ref{fig:bug_rootCause}). The reason is that the defined category focused on root causes related to APIs, models (structure and parameter), and documentation. Therefore, for further investigation of root causes, we went through the bugs categorized as \textit{Other} using the 11 labels provided by \cite{jia2021symptoms} for root causes of bugs in DL libraries. We followed a similar methodology to what is done initially for root causes. We found the following results: Dimension mismatch (1.3\%), Processing (15.2\%), Algorithm (15.2\%), Corner case (2.5\%), Logic error (36.7\%), Configuration error (19.0\%), Concurrency (10.1\%) while we found no occurrence (0\%) for Type confusion, Inconsistency, Referenced types error, and Memory. We refer the reader to \cite{jia2021symptoms} for a full description of the categories. The root cause of the majority of bugs is identified as Logic error, as it covers bugs that occur in the logic of  the code like incorrect program flow or wrong order of actions. This also includes bugs that lead to wrong calculations of gradients. Since ML code is highly dependent on configurations (like hyperparameters), the next largest portion of root causes is  Configuration error which represents bugs caused by wrong configurations. However, both classifications overlooked problems raised during training and using trained models. It looks like that, by increasing the development and usage of ML-based software systems, we need a new category of bug’s root causes for such systems.


Categorizing bugs based on their symptoms, as reported in Figure \ref{fig:bug_symptoms}, shows that crash with 52.2\% is the most prominent symptom which is in accordance with previous studies \cite{islam2019comprehensive}. However, the ratio of `incorrect functionality' rises to 39.4\% compared to 12\% in the previous report \cite{islam2019comprehensive}. The reason can be an increase in the application of ML components in software systems from 2019 (the previous study \cite{islam2019comprehensive}) to 2022.

\begin{figure}
  \centering
  \subfloat[]{\includegraphics[width=0.40\textwidth]{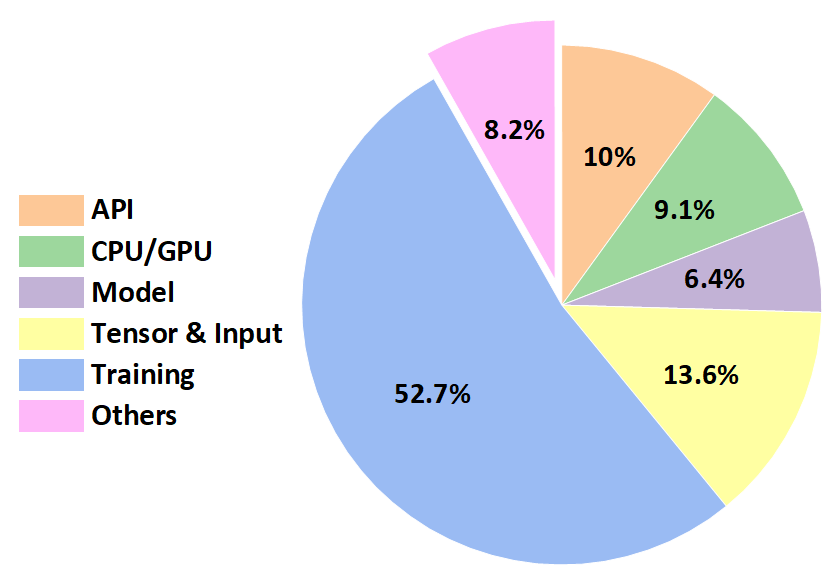}\label{fig:ml_bug_taxonomy}}
  \subfloat[]{\includegraphics[width=0.40\textwidth]{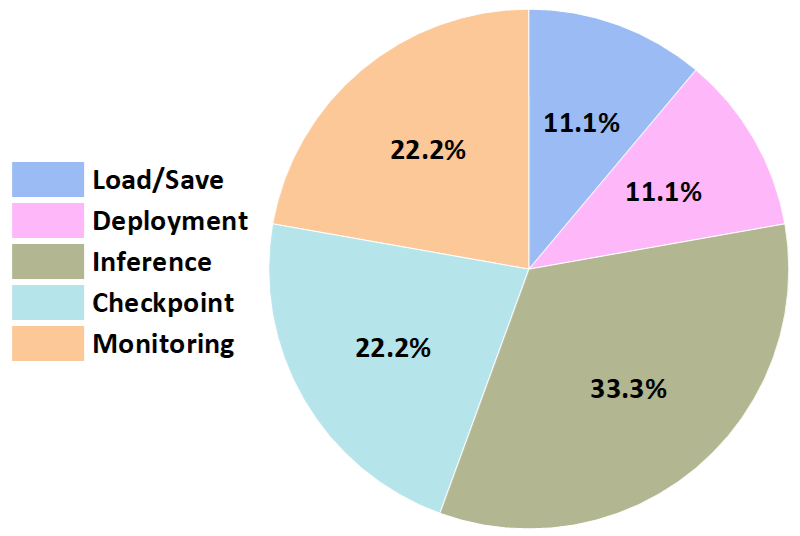}\label{fig:ml_bug_other}}
  
  \caption{Classification of bugs in ML-based systems (a) based on the taxonomy provided by \cite{humbatova2020taxonomy}, (b) bugs fall into `other' category.
  }
  \label{fig:bug_taxonomy}
  \vspace{-1em}
\end{figure}

Then, we categorized ML bugs based on the taxonomy of faults in DL systems introduced by Humbatova et al.~\cite{humbatova2020taxonomy}. Their taxonomy consists of 5 top-level categories that we used to label ML bugs: “Tensors\&Inputs”, “Model”, “Training”, “API”, and “GPU Usage”. The results are reported in Figure~\ref{fig:ml_bug_taxonomy}. Overall, the “Training” category is the most frequent type of bug with 52.7\% which is similar to the original study~\cite{humbatova2020taxonomy} where authors reported 52.5\% of issues related to the training. It also echoes~\cite{Long22} where accuracy/performance bugs are mainly concerned with the training phase (although it concerns bugs inside DL frameworks). The next prevalent type of bug is Tensors\&Inputs which deals with problems related to the wrong shape and type/format of the data. Similar to root causes, here, we encountered 9 bugs that did not fit into any of the 5 categories. So, we have added the ‘other’ category to cover them. These bugs include issues happening during the loading/saving of a trained model, deployment of the trained model, model inference, checkpoints, and monitoring of a trained model. Figure~\ref{fig:ml_bug_other} shows the distribution of bugs in the “other” category. The “loading/saving of a trained model” category refers to the issues which may occur when developers try to save a trained model to be able to use it in another program or load a saved model which is trained in the previous steps (e.g., issue~\cite{issue_sample_04}). Model deployment issues are related to the possible problems in deploying a trained ML model on a specific platform.  For instance, issue~\cite{issue_sample_05} tries to fix an architectural problem of the model to be able to deploy on ONNX. Problems categorized as inference issues refer to the errors occurring in the prediction step of the ML component. Issue~\cite{issue_sample_06} is an example of an inference issue explaining a problem in the implementation of the prediction step of the trained model. Checkpoint issues are explained as the problems that developers faced while trying to save the ML model during its training process to be able to continue the training process from checkpoints, in case of any crash/stop in the model training process. Issue~\cite{issue_sample_07} is considered an example of the model checkpoint issue. Monitoring issues are related to the problems of monitoring a deployed model. 

\begin{tcolorbox}
\textbf{Finding 2. } Because of the increasing usage of ML in different domains and the increasing maturity of ML-based software systems over time, we need to revise existing ML bug classifications and add new possible categories to be able to classify root causes/symptoms of ML bugs. Moreover, most of the issues of ML components seem to originate from the training phase.
\end{tcolorbox}

\subsection{RQ2: Complexity of bug fixes in ML-based systems}\label{sec:rq2}
\begin{figure*}
  \centering
  \subfloat[]{\includegraphics[width=0.40\textwidth]{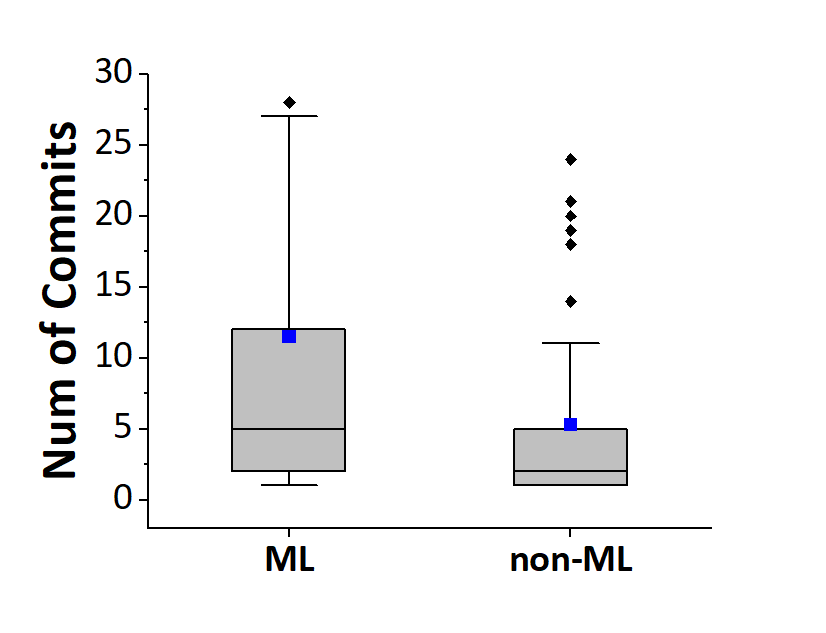}\label{fig:commits}}
  \subfloat[]{\includegraphics[width=0.40\textwidth]{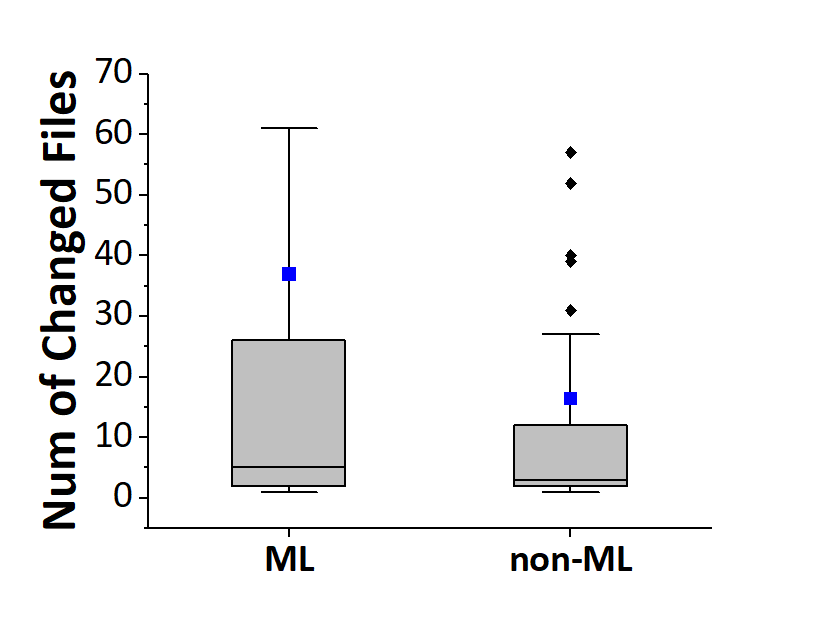}\label{fig:changed_files}}
  
  \subfloat[]{\includegraphics[width=0.40\textwidth]{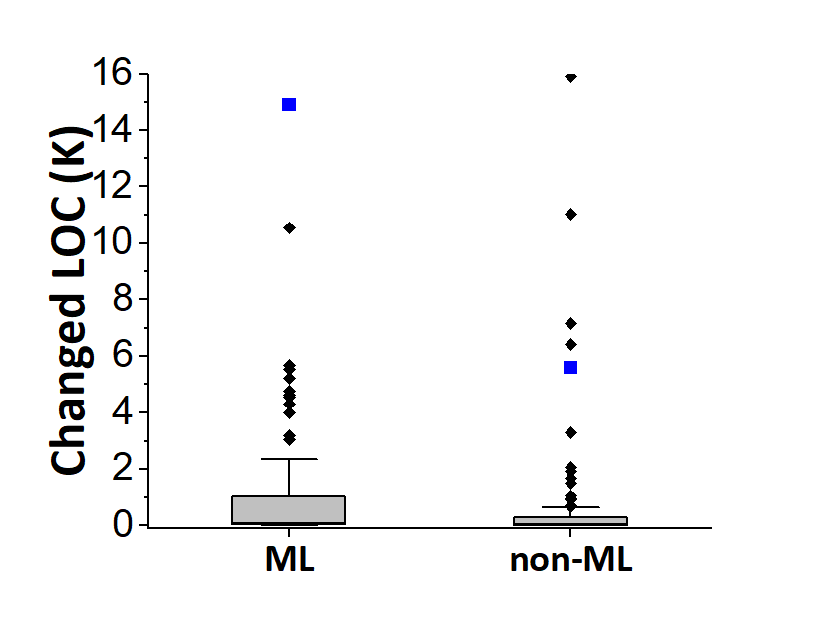}\label{fig:LOC}}
  \subfloat[]{\includegraphics[width=0.40\textwidth]{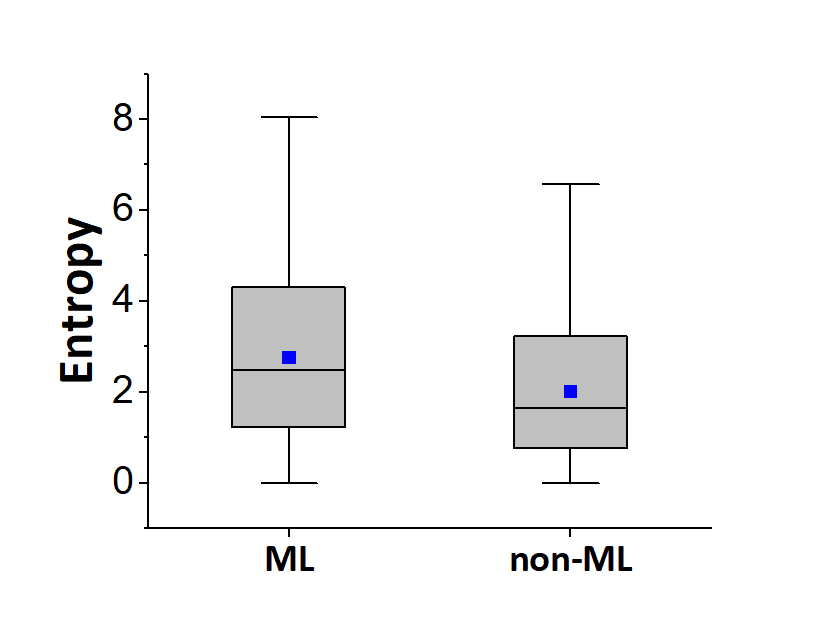}\label{fig:entropy}}
  
  \subfloat[]{\includegraphics[width=0.40\textwidth]{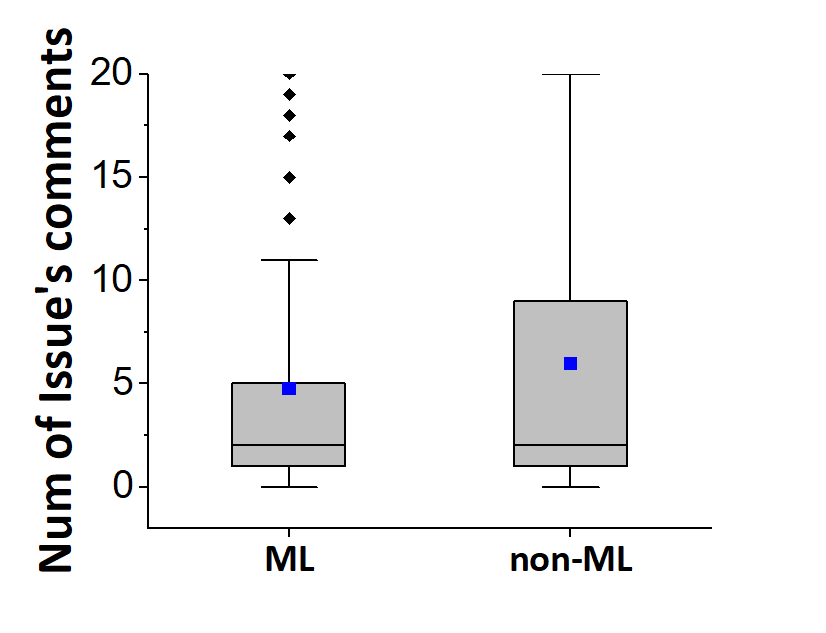}\label{fig:issue_comment}}
  \subfloat[]{\includegraphics[width=0.40\textwidth]{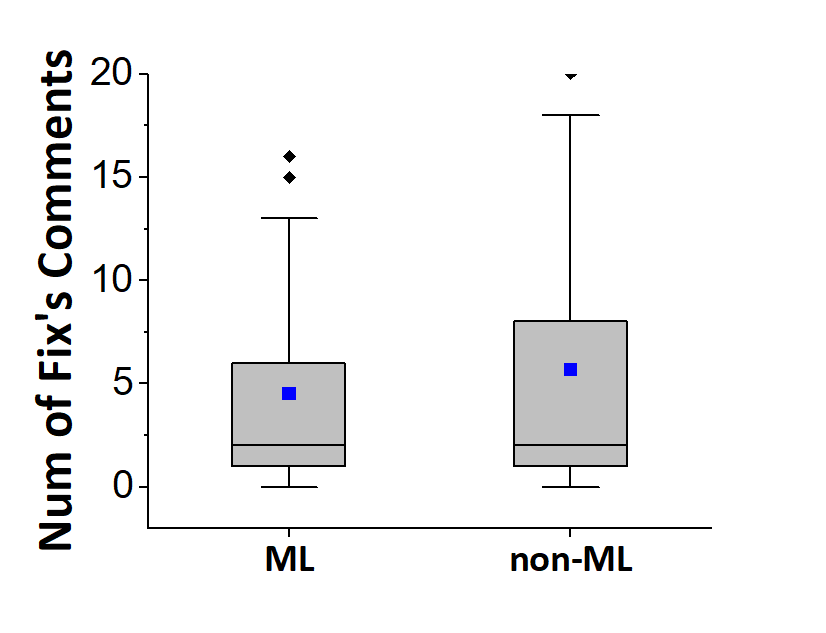}\label{fig:fix_comment}}
  
  \subfloat[]{\includegraphics[width=0.40\textwidth]{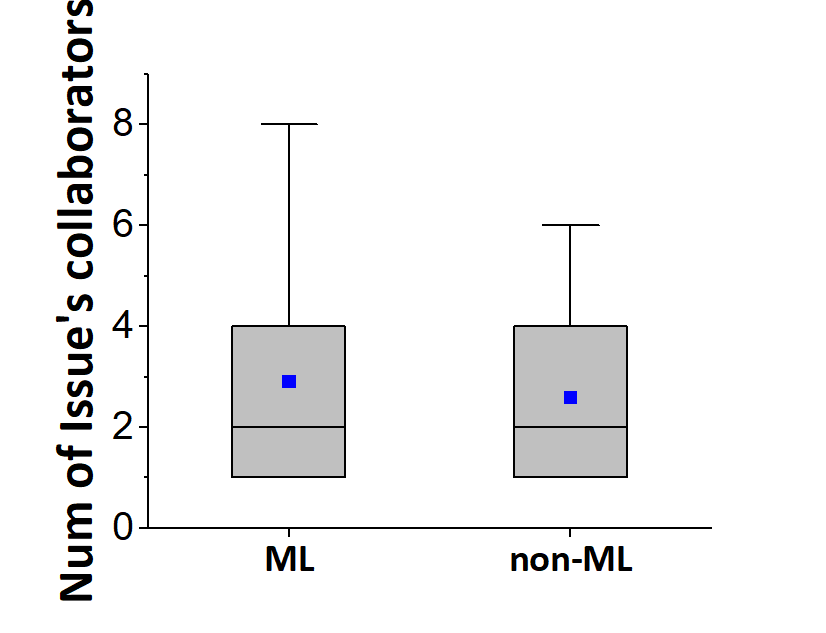}\label{fig:issue_collab}}
  \subfloat[]{\includegraphics[width=0.40\textwidth]{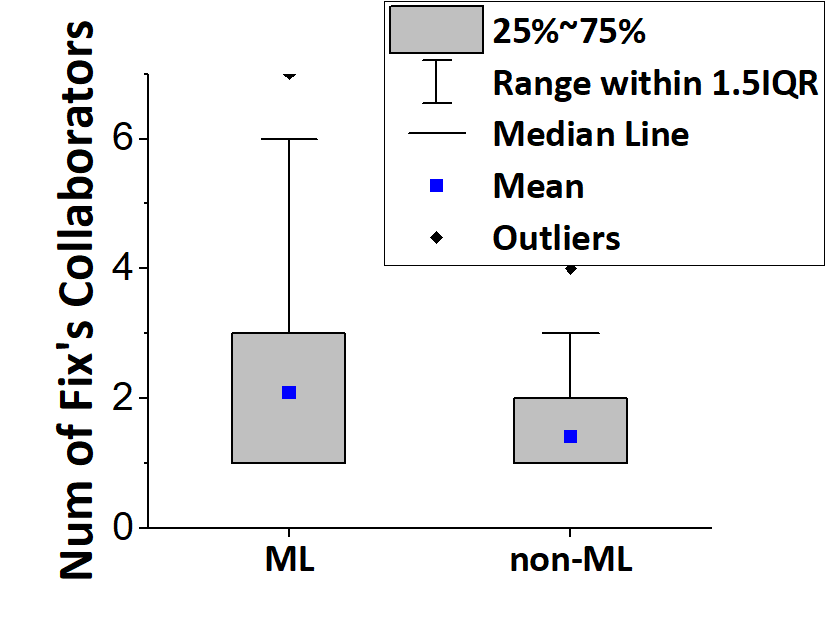}\label{fig:fix_collab}}
  
  \subfloat[]{\includegraphics[width=0.40\textwidth]{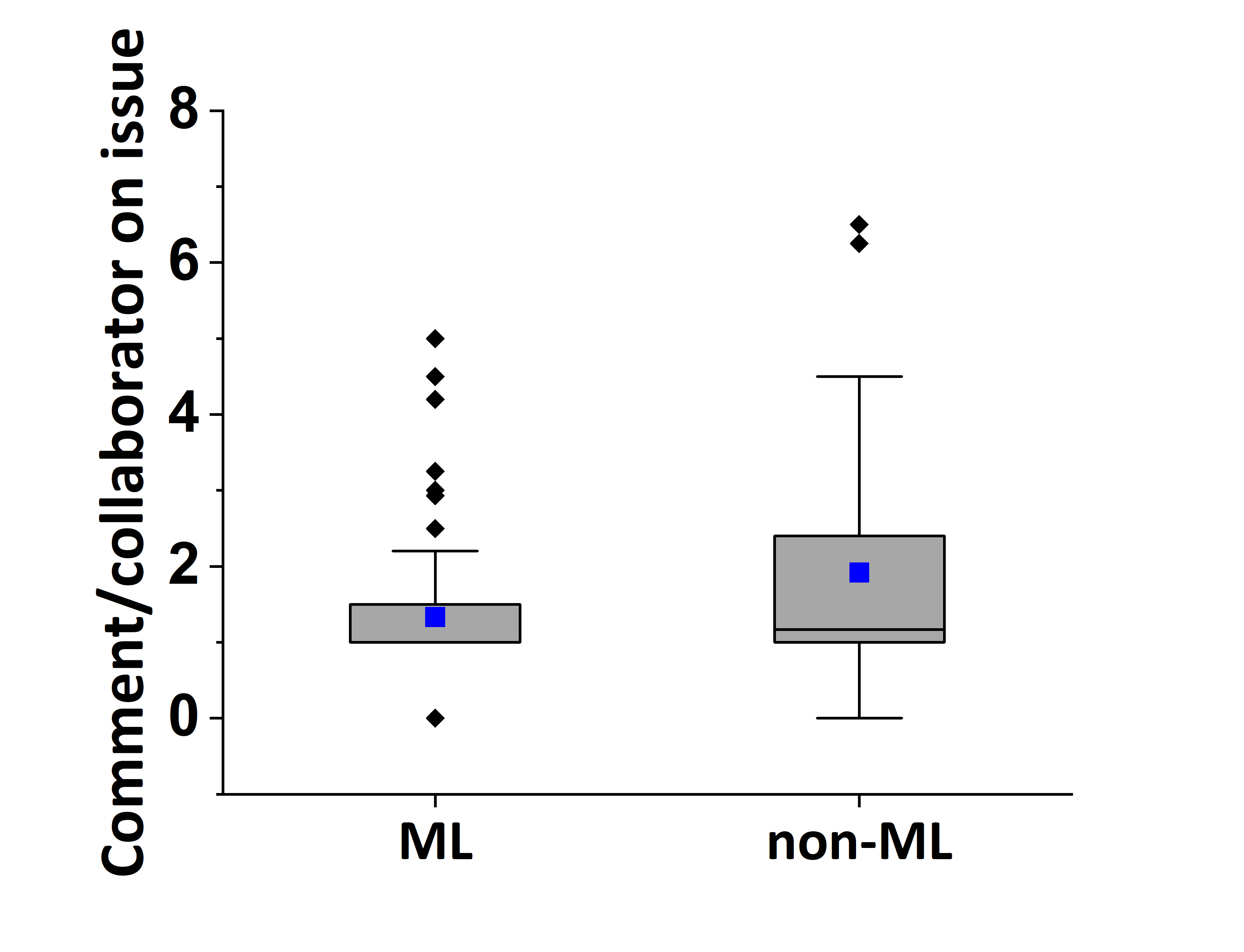}\label{fig:issue_comment_per_collab}}
  \subfloat[]{\includegraphics[width=0.40\textwidth]{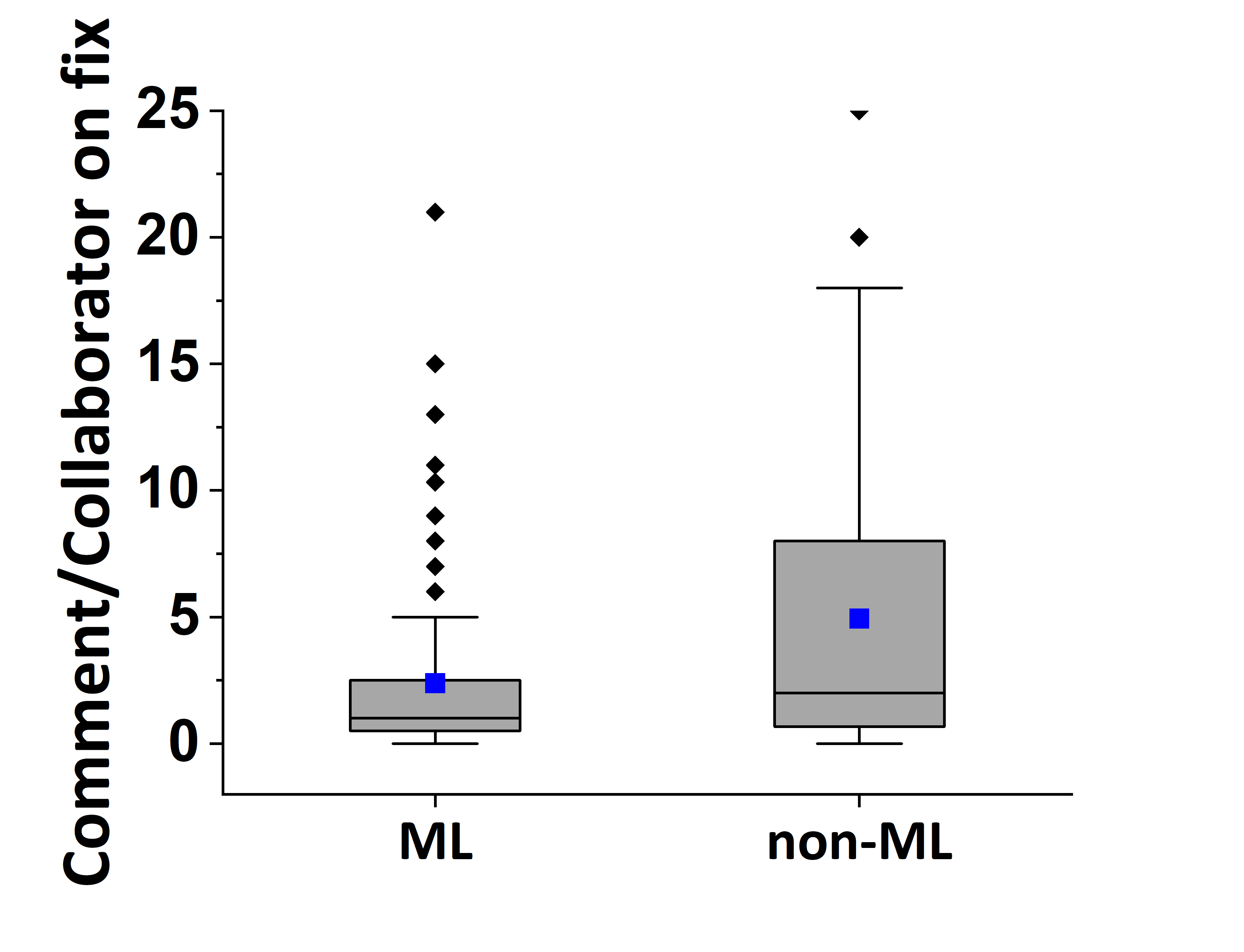}\label{fig:fix_comment_per_collab}}
  
  \caption{Complexity of Fixes: (a) number of bug-fixes commit, (b) number of bug-fixes changed files, (c) number of bug-fixes changed LOC, (d) bug-fixes entropy, (e) number of comments on an issue, (f) number of comments on bug-fix, (g) number of collaborators on an issue, and (h) number of collaborators on bug-fix, (i) number of comments/collaborator on issue, (j) number of comments/collaborator on a fix.
  }
  \label{fig:fix_complexity_level}
  \vspace{-1em}
\end{figure*}

To compare the complexity of bug fixes for different types of bugs in ML-based systems, we used metrics that are utilized successfully by previous studies~\cite{kononenko2018studying,chaturvedi2014predicting} including the number of commits to fix the bug, number of changed files, number of changed Lines of Code (LOC), entropy of changes, number of comments on the issue, number of comments on the bug-fix related to the issue, number of users who discussed the issue, and number of users who discussed and collaborated on bug-fix. Figure~\ref{fig:fix_complexity_level} reports the results. 

To ensure a statistically significant difference between metrics for ML and non-ML bugs, we also run statistical tests. As we make no assumptions on the distribution of data and since the variance of data is different, “Mann–Whitney U test” is the recommended test~\cite{arcuri2014hitchhiker}. We used the two-sided test of the \textit{Mann–Whitney U} test which tests the null hypothesis that the probability of a given metric of ML bugs being greater than non-ML bugs is equal to the probability of a given metric of non-ML bugs being greater than ML bugs. Finally, as reporting simply p-value might be misleading~\cite{Kampenes07}, we also computed effect size to assess the magnitude of the differences~\cite{Cliff93}. Moreover, “Cliff delta” test is a recommended test for calculating effect size for data with non-normal distribution~\cite{macbeth2011cliff}. For both tests, we used their official implementation in the R statistical environment (with “effsize” library for Cliff's delta). Results are presented in Table~\ref{tbl:statistical_test}.
 

\begin{table*}
\centering
\caption{Detailed result of \textit{Mann–Whitney U} and \textit{Cliff's delta} test.}
\begin{threeparttable}
   \begin{tabular}{p{5cm} r r}
        \hline
            \textbf{Metric} & \multicolumn{1}{c}{\textbf{\textit{p-value}}} & 
            \textbf{\textit{effect size (d)}}\\
            \hline
             \# commits & \textbf{$<$ 1e-3} & \textbf{0.337}\tnote{M}\\
             \# changed files & 0.080 & 0.133\tnote{N}\\
             \# changed LOC & \textbf{0.033} & \textbf{0.164}\tnote{S}\\
             \# comments on issue & 0.347 & -0.072\tnote{N}\\
             \# comments on fix & 0.441 & -0.058\tnote{N}\\
             \# collaborators on issue & 0.904 & -0.009\tnote{N}\\
             \# collaborator on fix & \textbf{$<$ 1e-3} & \textbf{0.272}\tnote{S}\\
             \# comment/collaborator on issue & 0.086 & -0.131\tnote{N}\\
             \# comment/collaborator on fix & \textbf{0.014} & \textbf{-0.192}\tnote{S}\\
             entropy & \textbf{0.008} & \textbf{0.204}\tnote{S}\\
        \hline
    \end{tabular}
    \begin{tablenotes}
       \item [M] medium effect 
       \item [S] small effect
       \item [N] negligible effect
     \end{tablenotes}
\end{threeparttable}
\label{tbl:statistical_test}
\end{table*}

As we can see in Table \ref{tbl:statistical_test}, regarding $p$-value, using the widely accepted $0.05$ significance threshold, the results show that differences in the number of commits, number of changed LOC, entropy of changes, number of collaboration on fix, 
and number of comment/collaborator on fixes
are significant while the number of changed files, number of collaborators on issues, number of comments on both issues and fixes, 
and number of comment/collaborator on issues
are not significant. To interpret the effect size, we use the traditional scale associated with Cliff's delta~\cite{Romano06} that is $\lvert d \rvert = 0.147$ is small, $\lvert d \rvert = 0.33$ is medium and $\lvert d \rvert = 0.474$ is large. All \textit{statistical} significant tests previously described are also \textit{practically} significant with a \textit{small} effect size, except \textit{number of commits} which results in a \textit{medium} effect size. Moreover, the number of commits, changed LOC, number of collaborators on a fix and entropy features have a positive effect size which indicates that this feature is more prevalent in ML bugs than non-ML bugs. 
Besides, the negative effect size of the number of comments/collaborators signifies that this feature tends to be higher in non-ML bugs compared to ML ones.
According to obtained results, ML bugs tend to require more commits as well as more changed LOC compared to non-ML bugs. They also require more collaborators to be fixed as well. Yet, the presence of numerous outliers in commits for non-ML bugs (Figure~\ref{fig:commits}) shows that some non-ML bugs can prove harder to address. Regarding the number of changed LOC, there seem to be multiple outliers in both cases (Figure~\ref{fig:LOC}), which can imply that this feature is independent of the nature of the bug, and is basically bug-dependent, although there is a significant difference.

To give a picture of the number of comments on issues and fixes, we provide the average number of comments per collaborator (Figures~\ref{fig:issue_comment_per_collab} and \ref{fig:fix_comment_per_collab}). Results reveal that collaborators put more comments on average on non-ML issues, compared to ML ones. As Table~\ref{tbl:statistical_test} represents, the difference between the number of comments/collaborators on issues in ML and non-ML bugs is not significant, but it is significant for fixes with small effect size ( $\lvert d \rvert = -0.192$).
One possible reason behind this discrepancy may be the fact that developers who put comments on ML bugs collaborate passively. In other words, collaborators just put a few comments on fixes and then stop collaborating on fixes actively, which may be because of their limited knowledge of ML. 
For instance, there are 8 different collaborators on a PR (\href{https://github.com/Lightning-AI/lightning/pull/4309}{\#4309}) of \textit{lightning-AI} project which is the fix of an issue (\href{https://github.com/Lightning-AI/lightning/issues/3660}{\#3660}). However, according to activities, only 2-3 of them are collaborating on fixing the issue. 
 
For the number of issue's collaborators (Figure~\ref{fig:issue_collab}), there is no statistical difference in our case which signifies that while ML bugs require more developers to be fixed (Figure~\ref{fig:fix_collab}), there is a similar number of collaborators discussing on the issue. As such, ML bugs tend not to require significantly more collaborators (experts) to be identified and diagnosed in ML-based systems. Regarding the number of comments on different issues (Figure~\ref{fig:issue_comment} and Figure~\ref{fig:fix_comment}), while not being statistically significant, the distribution of data reveals that the number of comments in non-ML bugs is on average higher than ML ones. The reason behind it can be that a higher number of developers are more skilled in general programming than in ML. However, there seems to be a high number of outliers, which means that while on average ML bugs receive fewer comments than non-ML ones, some bugs require extensive discussions, potentially harder to pinpoint the root cause of the problem for instance.
Finally, ML bugs tend to significantly have higher entropy than non-ML bugs (Figure~\ref{fig:entropy}).
This means, in general, ML bugs happen in some specific files whereas non-ML bugs tend to be more scattered across the project’s files. 

To further investigate why ML bugs need more change iterations (i.e., more commits), we performed a thematic analysis of the bug-fixing PRs 
using Latent Dirichlet Allocation (LDA) from the Gensim library \cite{lda-gensim}. We removed any kind of text known as PR template (repetitive text in all PRs)
which could bias the model in our 109 ML bugs' PRs. We then applied LDA for topic modeling based on Mallet \cite{mallet} to obtain the relevant number of topics. The algorithm returned the best coherence score (in parenthesis) for a number of topics of 7 (0.6326), 26 (0.6297), and 20 (0.6218). We chose the case of the 7 topics, as it yielded the best score and a higher number of topics might mean the model started grouping issues based on the application rather than the proper cause of the issue (for instance, in the 20 topics proposed, one of them is based on the keywords \enquote{state}, \enquote{action}, \enquote{intent} or \enquote{slot} which are keywords heavily used in one of the applications we collected). Reviewers then analyzed the keywords of the 7 topics and the issues labeled with the given topic to come up with topic labels. The final topics are: Test related (15.8\%), Requirements and Dependencies (19.3\%), Training Accelerator and Model checkpoint (18.4\%), Scripting Bugs (14.9\%), Model Bugs (16.7\%), Training Bugs (7.9\%), Documentation (7.0\%). The topics that are the most represented are commits linked to Requirements and Dependencies and Training Accelerator and Model checkpoint with over \textbf{18\%} each, while the least represented topics are the Documentation and Training bugs with around \textbf{7\%}. Interestingly, the largest part of the issues for ML bugs do not relate directly to ML (Requirements/Dependencies), nonetheless, such bugs are expected in ML systems as they often rely on multiple libraries and frameworks to work.

\begin{tcolorbox}
\textbf{Finding 3.} ML bugs have significantly different characteristics from non-ML ones, in ML-based systems: they need more commits, a higher number of LOC, and more collaborators involved. Analysis of the ML-related commits shows bugs based on Requirements/Dependencies are the most widespread while Documentation is the least represented.
\end{tcolorbox}

\subsection{RQ3: Needed resources for fixing bugs in ML-based systems}\label{sec:rq3}

To measure the number of resources spent for fixing bugs in ML-based systems, we consider two metrics: time-to-fix and expertise level of developers who fix the bugs (which were used in previous studies as well~\cite{kononenko2018studying,romano2021empirical}). 
To indicate the time-to-fix, we used the time range between the issue opening and closing date~\cite{bosu2014impact}. For the expertise level of developers, we consider the number of commits done in all other repositories by the developer who fixed the bug, exactly before fixing each issue of our dataset. Therefore, in our list, some users appeared multiple times with different numbers of commits depending on the date of the bug-fix. We kept only the highest expertise level observed (that is, the latest update of the number of commits). We added all users gathered that way whether they only were present on ML, non-ML, or both types of bugs in our dataset.
Similar to RQ2 (Section \ref{sec:rq2}), we use \textit{Mann-Whitney U} and \textit{Cliff's delta} test. 

Our results reveal that fixing time is not statistically significant ($p$-value = 0.423, d = 0.061 which is negligible), that is fixing ML bugs requires no more time compared to non-ML ones. 
This does not necessarily mean that ML bugs are as "easy to fix” as non-ML ones since other metrics we calculated such as the number of commits or the number of collaborators working on a fix were significantly higher for ML bugs. As we focused on repositories containing ML components, this non-significant difference might mean that developers are experts in solving both types of issues, so all issues take almost the same amount of time. This observation can be further emphasized when comparing the expertise level of users.

In all cases, the tests are not statistically significant which means that the number of prior commits of developers working on ML or non-ML issues is similar. This reflects on the observation made previously: as developers working in those repositories are used to both ML and non-ML issues, they are likely to be experts at the same level, even if in our dataset we flagged them as working on only one type of bug. Yet, some users worked on both types of issues. To assess the impact they have, we removed them from ML/non-ML issues groups and retested our data again, which resulted in still no significance even if the metric decreased. This means that users having some expertise in both types of bugs, in our dataset, do not change the outcome, implying that other users working on a single type of issue take as much time as the other group. Similarly, removing outliers (that is, users that have an overly high number of commits) does not affect the results. 

Finding 3 seems to suggest that ML bugs are more complex in terms of bug fixing than non-ML bugs. Nonetheless, the fact that we find no evidence of statistically different expertise levels or time-to-fix between ML and non-ML bugs in this study would seem to contradict Finding 3. As we do not have access to the \textit{effective time} spent on the fix, it can be the case that ML bugs were given priority and the actual time spent on fixing the non-ML issues is lower than the time-to-fix we have access to (date of fix and date of beginning). 
While there might be some threats to the validity of considered criteria used to measure needed resources for fixing ML and non-ML bugs (expertise level of developers and time-to-fix),
no other information was available to compute such a metric.

We examined the correlation between entropy and time-to-fix using the \textit{Spearman} correlation method from \textit{pandas} Python library.
Results reveal that time-to-fix has a moderate positive correlation (based on correlation formal definition~\cite{schober2018correlation}) with entropy (0.42). With respect to the fact that entropy is considered a measure to show the complexity of changes~\cite{chaturvedi2014predicting}, it can imply that time-to-fix is a good candidate to show the needed effort for fixing issues. 

Because of the negative impact of bug reopening on software quality \cite{tagra2022revisiting}, we examined the reopening rate of ML and non-ML bugs from the studied system. We aim to understand if developers experience higher bug reopening rates when dealing with ML bugs than they do for non-ML bugs. Results show that the reopening rate of ML issues (5.45\%) is a bit higher than that of non-ML bugs (4.2\%). We assessed the statistical significance of this result using the \textit{Chi-squared} test. We leveraged the \textit{scipy} Python library to run this test and obtained a $p-$value of $0.894$. Meaning that there is no significant difference between the reopening rate of ML and non-ML bugs. Besides, the rate of issue reopening in ML-based systems with 4.8\% is a bit less than the one reported for traditional software systems
(between 6\% and 10\%) by Zimmermann et al. ~\cite{zimmermann2012characterizing}.

Considering the major impact of code reviews on the software quality~\cite{bosu2016process}, we examined the rate of developers' invitations to review PRs fixing ML and non-ML bugs. We found that in 49.05\% of ML bugs, developers were invited to review PRs compared to only  23.36\% for non-ML-bugs. This finding suggests that higher attention is given to the quality of bug-fix changes when dealing with ML bugs. We explain this result by the fact that fixing ML bugs requires ML knowledge which may not be mastered by all of the developers on the team. We also investigated the number of PRs' reviewers in ML and non-ML issues since it has a positive correlation with the time-to-fix~\cite{maddila2019predicting}. As is shown in Figure~\ref{fig:pr_review}, the number of developers who reviewed bug-fixing PRs for ML bugs is higher than the number of developers who reviewed bug-fixing PRs for non-ML bugs. Furthermore, results of \textit{Mann–Whitney U} and \textit{Cliff's delta} tests (similar to \ref{sec:rq2}) indicate that the difference between the number of reviewers involved in fixing ML and non-ML issues is significant ($p$-value $<$ 1e-6, d = 0.378 which is interpreted as medium effect size).

Moreover, we checked the number of bug-fix PRs merged without review in our collected bugs. We found that 37.8\% of merging bug-fix PRs for non-ML bugs have been accomplished without conducting any review compared to 13.7\% for ML bugs. This lower number of non-reviewed bug-fixing PRs may suggest that development teams are more cautious when dealing with PRs related to ML bugs (reviewing 87.3\% of them), which may also explain the lower rate of bug reopening observed for ML bugs. 

\textcolor{blue}{
\begin{figure}
    \centering
    \includegraphics[width=0.5\textwidth]{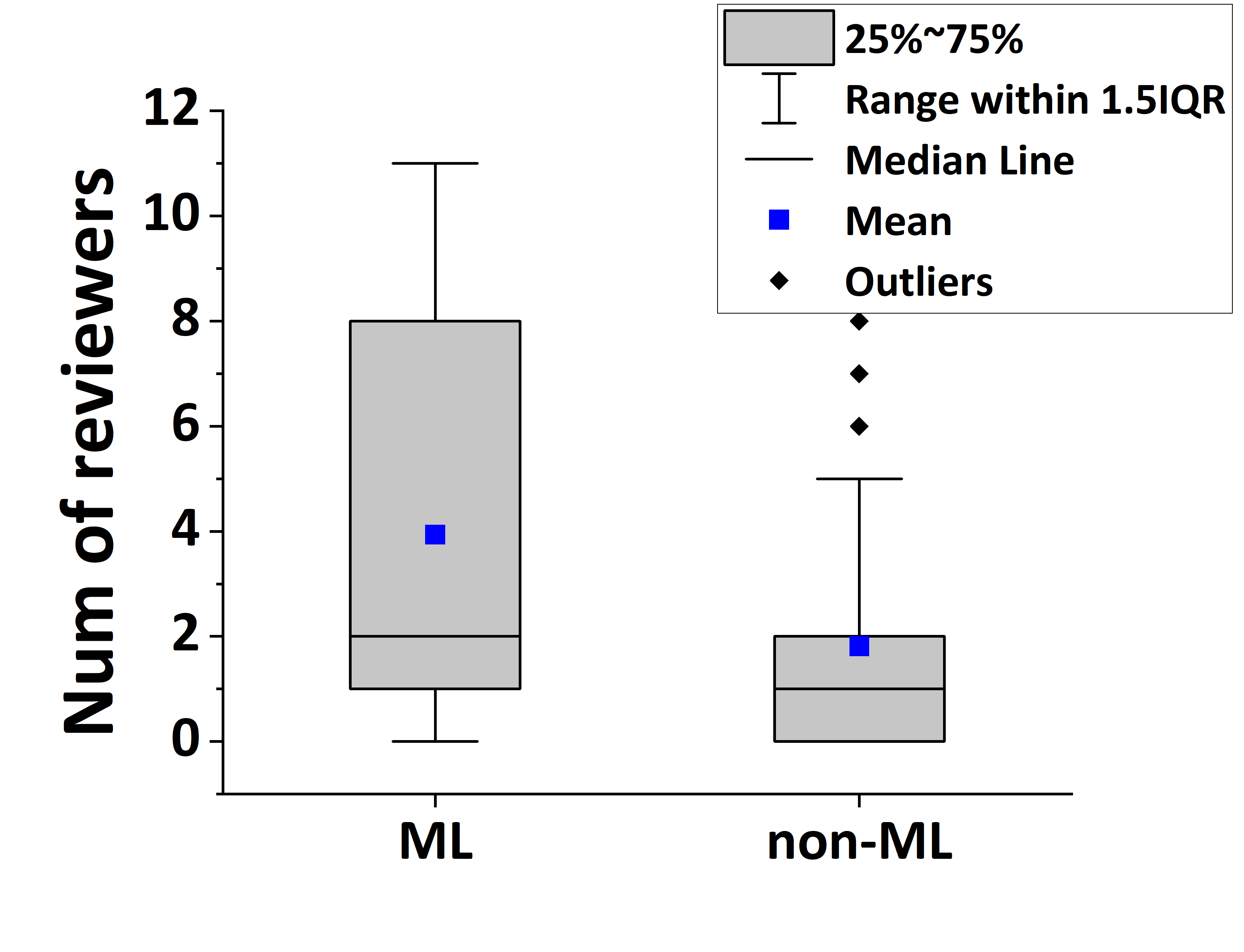}
    \caption{Number of developers who review PRs that fixed ML and non-ML bugs}
    \label{fig:pr_review}
\end{figure}
}



\begin{tcolorbox}
\textbf{Finding 4.} There is no statistically significant difference between ML and non-ML  on expertise level and time-to-fix when fixing bugs in ML-based systems. However, since the time-to-fix an issue does not capture all the effort that went into fixing the issue because it doesn't necessarily account for preparatory work done before starting editing the code, future work should investigate the cost of fixing ML bugs in more detail. 
\end{tcolorbox}



\subsection{Discussions}
We found that a large number of issues in ML-based systems either are users’ questions/mistakes or do not come with enough information to decide whether it is a bug or to help fix it. Moreover, although almost 10\% of development time is spent on ML components, nearly half of the “real” bugs in ML-based systems are related to ML components. Given the increasing usage and maturity of ML-based software systems over time, we believe that the software engineering community needs to revise existing ML bug classifications and add new possible categories to be able to classify root causes/symptoms of ML bugs. Our results revealed that ML bugs have significantly different characteristics from non-ML bugs, in ML-based systems: they need more number of commits, a higher number of LOC, and more collaborators involved, i.e. they seem to need more revision on average before the bug is fixed. The analysis of ML-related commits shows that bugs related to Requirements/Dependencies are the most widespread while Documentation bugs are the least represented. We did not find any statistically significant difference between ML and non-ML on the level of expertise of developers involved in their correction and the time taken to fix them in ML-based systems. However, the time-to-fix of an issue does not capture all the effort that went into fixing the issue because it does not necessarily account for preparatory work done before starting to edit the code.

Our work can help in planning and forecasting quality assurance activities for ML-based systems. Based on our findings, the maintenance team can allocate appropriate resources for detection and correction of bugs. Our findings show where and how ML-based system developers should invest their efforts to have the most efficient software maintenance (lowering the cost of maintenance). Similar to previous research works (e.g., compiler bugs in DL systems~\cite{shen2021comprehensive}), we shed light on the root causes and symptoms of bugs in open-source ML-based systems. These findings can be used to facilitate the development of (automatic) debugging tools for ML-based software systems~\cite{tan2014bug}. Furthermore, our findings indicate the necessity of adopting novel techniques for the prediction and detection of ML bugs, given their distinctive characteristics and distribution of ML bugs in comparison to non-ML bugs, as shown in this study.

It should be taken into consideration that applying ML in software systems results in generating new types of bugs which do not exist in traditional software systems. On the other hand, although developing ML components take just about 10\% of the ML-based systems development time~\cite{menzies2019five}, it represents roughly half of the real issues that a typical ML-based system will face. That is, ML components are more error-prone causing more serious suffering in comparison with non-ML ones. From the software maintenance viewpoint, it is obvious that developing ML components places a heavier financial burden on the software development process, with respect to their higher proneness to the bugs. As a software maintenance task, this observation emphasizes the need for automatic testing tools for ML components and ML-based software systems. A number of studies on the automatic bug detection~\cite{nikanjam2021automatic,zhang2021autotrainer}, bug localization~\cite{wardat2021deeplocalize}, debugging~\cite{wardat2022deepdiagnosis}, and bug repair~\cite{islam2020repairing} in ML-based systems that have been carried out during the last few years showed that there is an increasing need for automatic testing tools for ML-based systems.

Based on our dataset of ML bugs, one can develop an automatic approach to predict the type of bugs in ML-based systems, as the most recent taxonomy~\cite{humbatova2020taxonomy} does not cover deployment issues. By increasing development and usage of ML-based software systems in various new areas, developers may face new categories of bugs which raises demand for revising existing bug’s taxonomy and adding new discovered bugs to them. In other words, a novel taxonomy of bugs that covers issues related to all stages of development, deployment and maintenance of ML-based software systems is necessary. Zhang et al.~\cite{zhang2018empirical} is considered as one of the first studies that provided a list of root causes and symptoms of various bugs in ML-based systems. Shen et al.~\cite{shen2021comprehensive} also studied bugs related to ML-based systems from ML compiler point of view.  In fact, they studied bugs in ML compilers (such as TVM\footnote{https://tvm.apache.org/}, Glow\footnote{https://ai.meta.com/tools/glow/}, and nGraph\footnote{https://www.intel.ca/content/www/ca/en/artificial-intelligence/ngraph.html}), not in the ML-based systems. Tomban et al.~\cite{tambon2021silent} also generated a list of bugs which may occur inside ML frameworks. Therefore, it is obvious that understanding about bugs in ML-based systems is still progressing. Accordingly, we would be required to update ML bug’s root causes, based on the newly added categories of ML bugs. 

We believe there might be some specific metrics (compared to the studied metrics) to measure the complexity of ML components. There exist some studies reviewing complexity of the ML components from different viewpoints. As an example, Hu et al.~\cite{hu2021model} studied the complexity of DL models as the complexity of the problem that the DL model can express. They explained that model complexity can be measured based on various factors including model framework (feedforward neural network, convolutional neural network , etc), model size (number of parameters, number of hidden layers, etc), model optimization process (objective function’s form, learning algorithm, hyperparameters, etc), and the complexity of the data used for model training (data distribution, data dimensionality, information volume, etc). Zhang et al.~\cite{zhang2020machine} also represents model relevance as an ML testing property that checks the complexity of the model and tries to prevent model from overfitting. When the complexity of the model is higher than the problem, the model overfits to the training data and can not generalize. On the other hand, Yao et al.~\cite{yao2017complexity} carried out an empirical study to show the relationship between model complexity and model performance. They introduced complexity as the ability to control data preprocessing, feature selection, classifier selection and parameter tuning. Their results showed that models with higher complexity (more dimension to be controlled) gain better performance and classifier selection plays the most effective role in achieving higher performance. Therefore, ML engineers should be careful about the model complexity and the type of classifiers that they use.

With respect to the fact that about 70\% of software development costs belong to software maintenance~\cite{grubb2003software}, automatic bug detection, and debugging tools improve the quality and speed of software development leading to a decrease in the software maintenance cost and accordingly software development, significantly~\cite{bennett2000software}. 

\section{Related Works}
\label{sec:related_work}

Multiple studies analyzed bugs inside DL programs built on top of DL frameworks. Zhang et al.~\cite{zhang2018empirical} studied the characteristics of bugs in DL programs implemented based on \textit{TensorFlow}. They collected 175 bugs and reported their symptoms, root causes, and challenges to detect and localize them. Islam et al.~\cite{islam2019comprehensive} worked by studying bugs in DL programs implemented using \textit{Caffe}, \textit{Keras}, \textit{TensorFlow}, \textit{Theano}, \textit{Torch}. Humbatova et al.~\cite{humbatova2020taxonomy} manually investigated 1981 bug-fix commits and 1,392 issues/PRs from \textit{GitHub}, and 2,653 posts from \textit{Stack Overflow} related to the programs using \textit{TensorFlow}, \textit{Keras}, and \textit{PyTorch}, providing a taxonomy of bugs in DL programs by combining the result of manual checking and interviews of DL practitioners and researchers. Cao et al.\cite{cao2021characterizing} studied performance bugs in DL programs using \textit{TensorFlow} and \textit{Keras} from 225 \textit{Stack Overflow} posts. Finally, Liu et al.~\cite{liu2021detecting} performed an empirical study on the 12,289 failed \textit{TensorFlow} jobs and they provided a tool named \textit{ShapeTracer} to detect shape-related bugs in \textit{TensorFlow} programs. Although all of the mentioned studies conducted empirical studies on ML bugs, none of them studied the characteristics of ML bugs from the software maintenance perspective and the differences between ML and non-ML bugs in ML-based systems.  

Some other studies ~\cite{rivera2021challenge,tambon2021silent,jia2021symptoms} also investigated DL bugs, but inside the DL frameworks. For instance,
Jia et al. \cite{jia2021symptoms} studied root causes and symptoms of bugs affecting Tensorflow frameworks and provided a taxonomy. As such, since the bugs are not in the codes developed using those frameworks but rather in the framework itself, their study scope is different from ours.

\section{Threats to Validity}
\label{sec:validity}

\textit{Construct validity:} Limitations of our approach could come from: 1) keywords used for extracting repositories, and 2) choice of exclusion criteria for repositories and issues. The rest of the approach (bug filtering, issues classification of symptoms/root cause, statistical sampling...) is based on existing works. For keywords, we used `\texttt{import <ML framework>}' and `\texttt{language: python}' which is sufficiently general to catch a lot of repositories and is more likely to generate False Positive (repositories that we would end up discarding) rather than False Negative. Exclusion criteria were chosen after some preliminary discussions among authors to weed out irrelevant repositories. Raters further agreed to flag issues/repositories they were not sure about, to allow for careful discussion in order to avoid incorrect labeling.

\textit{Internal Validity:} The first source of internal validity is the manual checking of repositories and  issues. To alleviate this threat, the first three authors 
(one Ph.D. and two Ph.D. candidates who are practitioners of ML-based systems)
selected 300 repositories, and 386 issues, and then labeled them. After several meetings, we reached an agreement on the labeling process and the rules to differentiate between bug types. Another internal threat to validity is using sampling for gathering issues from the extracted repositories. To mitigate the sampling bias, we did the sampling with 95\% and 5\% of confidence levels and confidence intervals, respectively, which are the most common sampling in the Software Engineering community. The next threat is the fixes that we consider for each issue. In general, it is supposed that each PR works on one specific issue. But in some cases, developers use one PR to fix more than one issue. So, in a few cases, we could not separate the changes that are exactly related to the studied issues which may impact our results.

\textit{External Validity:} The main threat to external validity is the selected ML frameworks. We studied repositories using \textit{TensorFlow}, \textit{Keras}, and \textit{PyTorch}, because these frameworks are the most popular ML frameworks in GitHub, with the highest number of stars/forks among all ML frameworks (such as \textit{MXNet} \cite{chen2015mxnet}, and \textit{Caffe} \cite{jia2014caffe}). While we focused on \textit{Python} code inside ML repositories, as it is the most used language for this task \cite{voskoglou2017best,Gupta:MLLangugae}, we believe that similar observations could be made about ML issues for other programming languages, since the type of issue will likely remain the same. In the sampling process as well, we focus on repositories with the highest number of stars, forks, commits, and issues to study the most mature repositories and mitigate the generalization threat.
Another threat to the external validity of this study would be the generalization of the results, because of using GitHub as the main source for data collection. We selected GitHub because it is the most important platform for hosting open-source projects. 

\textit{Conclusion Validity:} Conclusion limitations can be potentially wrongly classified, missing issues/repository type, and the replicability of the study. We manually inspected 4,057 repositories out of over 400,000 repositories scraped from GitHub. Similarly, for issues, we extracted $44,342$ issues from the ML-based repositories and sampled $386$ using a well-known statistical procedure to ensure the credibility of our results. In both cases, we believe that the sample data is big enough to be representative, and not mislead us in our conclusions. Similarly, the labeling of both repositories and issues was done independently by three raters and then discussed to mitigate potential errors. At last, we provided a replication package \cite{Replication-Package} to allow for the reproducibility of our results, and also for other researchers to build on our study.

\section{Conclusion and Future Works}
\label{sec:conclusion}
Since ML-based systems are increasingly being used in various domains, including safety-critical systems. Their reliability has become paramount. Corrective maintenance is one of the main tasks in reliability engineering and bug diagnostic plays a key role in this task. In this paper, we aimed to characterize different types of bugs in ML-based systems, from the software maintenance perspective. Therefore, we manually checked 386 selected issues raised in ML-based systems developed using the three most popular ML frameworks (\textit{TensorFlow}, \textit{Keras}, and \textit{PyTorch}). Firstly, we showed that the prevalence of ML bugs is almost the same as non-ML bugs. Furthermore, the results showed that the previous classifications on ML-related bug types, their root cause, and symptoms are becoming stale and should be revised. We also observed that ML bugs are mostly more complicated than non-ML bugs. Moreover, the required resources (time-to-fix and developer expertise level) for fixing ML bugs are almost similar to non-ML ones. 
Besides, we represented that development teams fix ML issues 
more cautiously compared to
non-ML ones,
with respect to the percentage of bug-fix PRs which are merged without review. 
In our future works, our first objective is to illustrate the characteristics of the bugs in various types of ML algorithms such as Computer Vision, NLP, etc. Additionally, we aim to expand the current study to comprehensively characterize different types of ML bugs (and their root causes/symptoms) and make a comparison among the characterization of different types of ML bugs. Furthermore, we plan to revise the existing taxonomy of bug types to encompass issues encountered throughout the whole ML application development pipeline including development, deployment, and maintenance, not only pre-processing, training, and testing. Finally, following previous studies~\cite{hanam2016discovering} ML bug characteristics can be used as a base to identify bug patterns in ML-based systems which help ML software maintainers to detect/predict ML bugs more easily. 

\section{Data Availability Statement}
The dataset generated during the current study is available in the replication package, which is accessible via 
\cite{Replication-Package}.

\bibliographystyle{IEEEtran}
\bibliography{bibliography}

\begin{thebibliography}{100}
\providecommand{\url}[1]{#1}
\csname url@samestyle\endcsname
\providecommand{\newblock}{\relax}
\providecommand{\bibinfo}[2]{#2}
\providecommand{\BIBentrySTDinterwordspacing}{\spaceskip=0pt\relax}
\providecommand{\BIBentryALTinterwordstretchfactor}{4}
\providecommand{\BIBentryALTinterwordspacing}{\spaceskip=\fontdimen2\font plus
\BIBentryALTinterwordstretchfactor\fontdimen3\font minus
  \fontdimen4\font\relax}
\providecommand{\BIBforeignlanguage}[2]{{%
\expandafter\ifx\csname l@#1\endcsname\relax
\typeout{** WARNING: IEEEtran.bst: No hyphenation pattern has been}%
\typeout{** loaded for the language `#1'. Using the pattern for}%
\typeout{** the default language instead.}%
\else
\language=\csname l@#1\endcsname
\fi
#2}}
\providecommand{\BIBdecl}{\relax}
\BIBdecl

\bibitem{liu2020small}
Z.~Liu, D.~Li, S.~S. Ge, and F.~Tian, ``Small traffic sign detection from large
  image,'' \emph{Applied Intelligence}, vol.~50, no.~1, pp. 1--13, 2020.

\bibitem{aithal2021automatic}
S.~G. Aithal, A.~B. Rao, and S.~Singh, ``Automatic question-answer pairs
  generation and question similarity mechanism in question answering system,''
  \emph{Applied Intelligence}, vol.~51, no.~11, pp. 8484--8497, 2021.

\bibitem{morovati2023bugs}
M.~M. Morovati, A.~Nikanjam, F.~Khomh, and Z.~M. Jiang, ``Bugs in machine
  learning-based systems: a faultload benchmark,'' \emph{Empirical Software
  Engineering}, vol.~28, no.~3, p.~62, 2023.

\bibitem{ieee5733835:2010}
IEEE, \emph{ISO/IEC/IEEE International Standard - Systems and software
  engineering -- Vocabulary}.\hskip 1em plus 0.5em minus 0.4em\relax 3 Park
  Avenue, New York NY 10016-5997, USA: IEEE, 2010.

\bibitem{sculley2015hidden}
D.~Sculley, G.~Holt, D.~Golovin, E.~Davydov, T.~Phillips, D.~Ebner,
  V.~Chaudhary, M.~Young, J.-F. Crespo, and D.~Dennison, ``Hidden technical
  debt in machine learning systems,'' \emph{Advances in neural information
  processing systems}, vol.~28, 2015.

\bibitem{grubb2003software}
P.~Grubb and A.~A. Takang, \emph{Software maintenance: concepts and
  practice}.\hskip 1em plus 0.5em minus 0.4em\relax World Scientific, 2003.

\bibitem{zhang2020machine}
J.~M. Zhang, M.~Harman, L.~Ma, and Y.~Liu, ``Machine learning testing: Survey,
  landscapes and horizons,'' \emph{IEEE Transactions on Software Engineering},
  vol.~48, no.~01, pp. 1--36, jan 2022.

\bibitem{zhang2018empirical}
Y.~Zhang, Y.~Chen, S.-C. Cheung, Y.~Xiong, and L.~Zhang, ``An empirical study
  on tensorflow program bugs,'' in \emph{Proceedings of the 27th ACM SIGSOFT
  International Symposium on Software Testing and Analysis}.\hskip 1em plus
  0.5em minus 0.4em\relax 3 Park Avenue, New York NY 10016-5997, USA: IEEE,
  2018, pp. 129--140.

\bibitem{shen2021comprehensive}
\BIBentryALTinterwordspacing
Q.~Shen, H.~Ma, J.~Chen, Y.~Tian, S.-C. Cheung, and X.~Chen, ``A comprehensive
  study of deep learning compiler bugs,'' in \emph{Proceedings of the 29th ACM
  Joint Meeting on European Software Engineering Conference and Symposium on
  the Foundations of Software Engineering}, ser. ESEC/FSE 2021.\hskip 1em plus
  0.5em minus 0.4em\relax New York, NY, USA: Association for Computing
  Machinery, 2021, p. 968–980. [Online]. Available:
  \url{https://doi.org/10.1145/3468264.3468591}
\BIBentrySTDinterwordspacing

\bibitem{yan2021exposing}
\BIBentryALTinterwordspacing
M.~Yan, J.~Chen, X.~Zhang, L.~Tan, G.~Wang, and Z.~Wang, ``Exposing numerical
  bugs in deep learning via gradient back-propagation,'' in \emph{Proceedings
  of the 29th ACM Joint Meeting on European Software Engineering Conference and
  Symposium on the Foundations of Software Engineering}, ser. ESEC/FSE
  2021.\hskip 1em plus 0.5em minus 0.4em\relax New York, NY, USA: Association
  for Computing Machinery, 2021, p. 627–638. [Online]. Available:
  \url{https://doi.org/10.1145/3468264.3468612}
\BIBentrySTDinterwordspacing

\bibitem{humbatova2020taxonomy}
\BIBentryALTinterwordspacing
N.~Humbatova, G.~Jahangirova, G.~Bavota, V.~Riccio, A.~Stocco, and P.~Tonella,
  ``Taxonomy of real faults in deep learning systems,'' in \emph{Proceedings of
  the ACM/IEEE 42nd International Conference on Software Engineering}, ser.
  ICSE '20.\hskip 1em plus 0.5em minus 0.4em\relax New York, NY, USA:
  Association for Computing Machinery, 2020, p. 1110–1121. [Online].
  Available: \url{https://doi.org/10.1145/3377811.3380395}
\BIBentrySTDinterwordspacing

\bibitem{islam2019comprehensive}
\BIBentryALTinterwordspacing
M.~J. Islam, G.~Nguyen, R.~Pan, and H.~Rajan, ``A comprehensive study on deep
  learning bug characteristics,'' in \emph{Proceedings of the 2019 27th ACM
  Joint Meeting on European Software Engineering Conference and Symposium on
  the Foundations of Software Engineering}, ser. ESEC/FSE 2019.\hskip 1em plus
  0.5em minus 0.4em\relax New York, NY, USA: Association for Computing
  Machinery, 2019, p. 510–520. [Online]. Available:
  \url{https://doi.org/10.1145/3338906.3338955}
\BIBentrySTDinterwordspacing

\bibitem{abadi2016tensorflow}
M.~Abadi, P.~Barham, J.~Chen, Z.~Chen, A.~Davis, J.~Dean, M.~Devin,
  S.~Ghemawat, G.~Irving, M.~Isard \emph{et~al.}, ``Tensorflow: A system for
  large-scale machine learning,'' in \emph{12th $\{$USENIX$\}$ symposium on
  operating systems design and implementation ($\{$OSDI$\}$ 16)}.\hskip 1em
  plus 0.5em minus 0.4em\relax Savannah, GA, USA: USENIX, 2016, pp. 265--283.

\bibitem{chollet2018keras}
F.~Chollet \emph{et~al.}, ``Keras: The python deep learning library,''
  Michigan, United States, pp. ascl--1806, 2018.

\bibitem{paszke2019pytorch}
A.~Paszke, S.~Gross, F.~Massa, A.~Lerer, J.~Bradbury, G.~Chanan, T.~Killeen,
  Z.~Lin, N.~Gimelshein, L.~Antiga \emph{et~al.}, ``Pytorch: An imperative
  style, high-performance deep learning library,'' Ithaca, NY, United States,
  2019.

\bibitem{Replication-Package}
``Replication package,''
  \url{https://github.com/ML-Bugs-2022/Replication-Package}, 2023.

\bibitem{lau2018introduction}
\BIBentryALTinterwordspacing
K.-K. Lau and S.~di~Cola, \emph{An Introduction to Component-Based Software
  Development}.\hskip 1em plus 0.5em minus 0.4em\relax World Scientific
  Publishing Co Pte Ltd: World Scientific, 2017. [Online]. Available:
  \url{https://www.worldscientific.com/doi/abs/10.1142/10486}
\BIBentrySTDinterwordspacing

\bibitem{martinez2021software}
S.~Mart{\'\i}nez-Fern{\'a}ndez, J.~Bogner, X.~Franch, M.~Oriol, J.~Siebert,
  A.~Trendowicz, A.~M. Vollmer, and S.~Wagner, ``Software engineering for
  ai-based systems: A survey,'' 2021.

\bibitem{carta2021multi}
S.~Carta, A.~Corriga, A.~Ferreira, A.~S. Podda, and D.~R. Recupero, ``A
  multi-layer and multi-ensemble stock trader using deep learning and deep
  reinforcement learning,'' \emph{Applied Intelligence}, vol.~51, no.~2, pp.
  889--905, 2021.

\bibitem{humbatova2021deepcrime}
\BIBentryALTinterwordspacing
N.~Humbatova, G.~Jahangirova, and P.~Tonella, ``Deepcrime: Mutation testing of
  deep learning systems based on real faults,'' in \emph{Proceedings of the
  30th ACM SIGSOFT International Symposium on Software Testing and Analysis},
  ser. ISSTA 2021.\hskip 1em plus 0.5em minus 0.4em\relax New York, NY, USA:
  Association for Computing Machinery, 2021, p. 67–78. [Online]. Available:
  \url{https://doi.org/10.1145/3460319.3464825}
\BIBentrySTDinterwordspacing

\bibitem{chen2020comprehensive}
\BIBentryALTinterwordspacing
Z.~Chen, Y.~Cao, Y.~Liu, H.~Wang, T.~Xie, and X.~Liu, ``A comprehensive study
  on challenges in deploying deep learning based software,'' in
  \emph{Proceedings of the 28th ACM Joint Meeting on European Software
  Engineering Conference and Symposium on the Foundations of Software
  Engineering}, ser. ESEC/FSE 2020.\hskip 1em plus 0.5em minus 0.4em\relax New
  York, NY, USA: Association for Computing Machinery, 2020, p. 750–762.
  [Online]. Available: \url{https://doi.org/10.1145/3368089.3409759}
\BIBentrySTDinterwordspacing

\bibitem{vlasic2016self}
\BIBentryALTinterwordspacing
B.~Vlasic and N.~E. Boudette. (2016) Self-driving tesla was involved in fatal
  crash, us says. The New York Times Company New York, NY, USA. [Online].
  Available:
  \url{https://www.nytimes.com/2016/07/01/business/self-driving-tesla-fatal-crash-investigation.html}
\BIBentrySTDinterwordspacing

\bibitem{scikit-learn}
\BIBentryALTinterwordspacing
F.~Pedregosa, G.~Varoquaux, A.~Gramfort, V.~Michel, B.~Thirion, O.~Grisel,
  M.~Blondel, P.~Prettenhofer \emph{et~al.}, ``Scikit-learn: Machine learning
  in python,'' \emph{J. Mach. Learn. Res.}, vol.~12, p. 2825–2830, nov 2011.
  [Online]. Available: \url{http://scikit-learn.sourceforge.net}
\BIBentrySTDinterwordspacing

\bibitem{schoop2021umlaut}
\BIBentryALTinterwordspacing
E.~Schoop, F.~Huang, and B.~Hartmann, ``Umlaut: Debugging deep learning
  programs using program structure and model behavior,'' in \emph{Proceedings
  of the 2021 CHI Conference on Human Factors in Computing Systems}, ser. CHI
  '21.\hskip 1em plus 0.5em minus 0.4em\relax New York, NY, USA: Association
  for Computing Machinery, 2021. [Online]. Available:
  \url{https://doi.org/10.1145/3411764.3445538}
\BIBentrySTDinterwordspacing

\bibitem{galin2004software}
D.~Galin, \emph{Software quality assurance: from theory to
  implementation}.\hskip 1em plus 0.5em minus 0.4em\relax \url{pearson.com}:
  Pearson education, 2004.

\bibitem{riccio2020testing}
V.~Riccio, G.~Jahangirova, A.~Stocco, N.~Humbatova, M.~Weiss, and P.~Tonella,
  ``Testing machine learning based systems: a systematic mapping,''
  \emph{Empirical Software Engineering}, vol.~25, no.~6, pp. 5193--5254, 2020.

\bibitem{issue_sample_01}
``Write complete json log after training,''
  \url{https://github.com/snorkel-team/snorkel/pull/1445}, 2019.

\bibitem{issue_sample_02}
``Interactive learning has server error hosting on docker,''
  \url{https://github.com/RasaHQ/rasa/issues/4142}, 2019.

\bibitem{issue_sample_03}
``Pps: fix of ispixelhit,'' \url{https://github.com/cms-sw/cmssw/pull/35089},
  2021.

\bibitem{lyu2007software}
M.~R. Lyu, ``Software reliability engineering: A roadmap,'' in \emph{Future of
  Software Engineering (FOSE'07)}.\hskip 1em plus 0.5em minus 0.4em\relax 3
  Park Avenue, New York NY 10016-5997, USA: IEEE, 2007, pp. 153--170.

\bibitem{IEEE:reliability:7827907}
IEEE, \emph{IEEE Recommended Practice on Software Reliability}.\hskip 1em plus
  0.5em minus 0.4em\relax 3 Park Avenue, New York NY 10016-5997, USA: IEEE,
  2017.

\bibitem{islam2020repairing}
\BIBentryALTinterwordspacing
M.~J. Islam, R.~Pan, G.~Nguyen, and H.~Rajan, ``Repairing deep neural networks:
  Fix patterns and challenges,'' in \emph{Proceedings of the ACM/IEEE 42nd
  International Conference on Software Engineering}, ser. ICSE '20.\hskip 1em
  plus 0.5em minus 0.4em\relax New York, NY, USA: Association for Computing
  Machinery, 2020, p. 1135–1146. [Online]. Available:
  \url{https://doi.org/10.1145/3377811.3380378}
\BIBentrySTDinterwordspacing

\bibitem{lenarduzzi2021software}
V.~Lenarduzzi, F.~Lomio, S.~Moreschini, D.~Taibi, and D.~A. Tamburri,
  ``Software quality for ai: Where we are now?'' in \emph{Software Quality:
  Future Perspectives on Software Engineering Quality}, D.~Winkler, S.~Biffl,
  D.~Mendez, M.~Wimmer, and J.~Bergsmann, Eds.\hskip 1em plus 0.5em minus
  0.4em\relax Cham: Springer International Publishing, 2021, pp. 43--53.

\bibitem{wardat2021deeplocalize}
M.~Wardat, W.~Le, and H.~Rajan, ``Deeplocalize: Fault localization for deep
  neural networks,'' in \emph{2021 IEEE/ACM 43rd International Conference on
  Software Engineering (ICSE)}.\hskip 1em plus 0.5em minus 0.4em\relax 3 Park
  Avenue, New York NY 10016-5997, USA: IEEE, 2021, pp. 251--262.

\bibitem{wang2006reliability}
H.~Wang, H.~Pham \emph{et~al.}, \emph{Reliability and optimal
  maintenance}.\hskip 1em plus 0.5em minus 0.4em\relax Springer International
  Publishing: Springer, 2006, vol. 14197.

\bibitem{ni2020analyzing}
Z.~Ni, B.~Li, X.~Sun, T.~Chen, B.~Tang, and X.~Shi, ``Analyzing bug fix for
  automatic bug cause classification,'' \emph{Journal of Systems and Software},
  vol. 163, p. 110538, 2020.

\bibitem{github-website}
``{G}ithub,'' \url{https://github.com/}, 2022.

\bibitem{li2020exploratory}
S.~Li, Y.~Wu, Y.~Liu, D.~Wang, M.~Wen, Y.~Tao, Y.~Sui, and Y.~Liu, ``An
  exploratory study of bugs in extended reality applications on the web,'' in
  \emph{2020 IEEE 31st International Symposium on Software Reliability
  Engineering (ISSRE)}.\hskip 1em plus 0.5em minus 0.4em\relax 3 Park Avenue,
  New York NY 10016-5997, USA: IEEE, 2020, pp. 172--183.

\bibitem{github_api_v3}
G.~developer~guideline documentation, ``Github rest api,''
  \url{https://developer.github.com/v3/}, 2021, accessed: 2021-7-27.

\bibitem{voskoglou2017best}
\BIBentryALTinterwordspacing
C.~Voskoglou. (2017) What is the best programming language for machine
  learning. Towards Data Science. [Online]. Available:
  \url{https://towardsdatascience.com/what-is-the-best-programming-language-for-machine-learning-a745c156d6b7}
\BIBentrySTDinterwordspacing

\bibitem{Gupta:MLLangugae}
S.~Gupta, ``What is the best language for machine learning?''
  \url{https://www.springboard.com/blog/data-science/best-language-for-machine-learning},
  2021, accessed: 2021-10-06.

\bibitem{github_GraphQL_API}
GitHub, ``Github graphql api documentation,''
  \url{https://docs.github.com/en/graphql}, 2021, accessed: 2021-7-27.

\bibitem{krishna2018connection}
R.~Krishna, A.~Agrawal, A.~Rahman, A.~Sobran, and T.~Menzies, ``What is the
  connection between issues, bugs, and enhancements?'' in \emph{2018 IEEE/ACM
  40th International Conference on Software Engineering: Software Engineering
  in Practice Track (ICSE-SEIP)}.\hskip 1em plus 0.5em minus 0.4em\relax 3 Park
  Avenue, New York NY 10016-5997, USA: IEEE, 2018, pp. 306--315.

\bibitem{hata2021same}
\BIBentryALTinterwordspacing
H.~Hata, R.~G. Kula, T.~Ishio, and C.~Treude, ``Same file, different changes:
  The potential of meta-maintenance on github,'' in \emph{Proceedings of the
  43rd International Conference on Software Engineering}, IEEE.\hskip 1em plus
  0.5em minus 0.4em\relax 3 Park Avenue, New York NY 10016-5997, USA: IEEE
  Press, 2021, p. 773–784. [Online]. Available:
  \url{https://doi.org/10.1109/ICSE43902.2021.00076}
\BIBentrySTDinterwordspacing

\bibitem{github_repo_chinese}
HanLP, ``Hanlp: Han language processing,''
  \url{https://github.com/hankcs/HanLP}, 2021, accessed: 2021-11-01.

\bibitem{github_repo_training}
Tensorflow, \url{https://github.com/tensorflow/models}, 2021, accessed:
  2021-11-01.

\bibitem{github_repo_data_preprocess}
NVIDIA, ``Nvtabular,'' \url{https://github.com/NVIDIA-Merlin/NVTabular}, 2021,
  accessed: 2021-11-01.

\bibitem{github_repo_test_func}
ultralytics, ``Yolov3,'' \url{https://github.com/ultralytics/yolov3}, 2021,
  accessed: 2021-11-01.

\bibitem{github_repo_wrapper}
Tensorpack, \url{https://github.com/tensorpack/tensorpack}, 2021, accessed:
  2021-11-01.

\bibitem{seaman1999qualitative}
C.~B. Seaman, ``Qualitative methods in empirical studies of software
  engineering,'' \emph{IEEE Transactions on software engineering}, vol.~25,
  no.~4, pp. 557--572, 1999.

\bibitem{keras_doc}
Keras, ``Formal documentation of keras apis,''
  \url{https://keras.io/api/models/}, 2022.

\bibitem{falotico2015fleiss}
R.~Falotico and P.~Quatto, ``Fleiss’ kappa statistic without paradoxes,''
  \emph{Quality \& Quantity}, vol.~49, no.~2, pp. 463--470, 2015.

\bibitem{yang2022mining}
Y.~Yang, T.~He, Y.~Feng, S.~Liu, and B.~Xu, ``Mining python fix patterns via
  analyzing fine-grained source code changes,'' \emph{Empirical Software
  Engineering}, vol.~27, no.~2, pp. 1--37, 2022.

\bibitem{quach2021empirical}
S.~Quach, M.~Lamothe, Y.~Kamei, and W.~Shang, ``An empirical study on the use
  of szz for identifying inducing changes of non-functional bugs,''
  \emph{Empirical Software Engineering}, vol.~26, no.~4, pp. 1--25, 2021.

\bibitem{hartling2012validity}
L.~Hartling, M.~Hamm, A.~Milne, B.~Vandermeer, P.~L. Santaguida, M.~Ansari,
  A.~Tsertsvadze, S.~Hempel, P.~Shekelle, and D.~M. Dryden, ``Validity and
  inter-rater reliability testing of quality assessment instruments,'' 2012.

\bibitem{nikanjam2022faults}
A.~Nikanjam, M.~M. Morovati, F.~Khomh, and H.~Ben~Braiek, ``Faults in deep
  reinforcement learning programs: a taxonomy and a detection approach,''
  \emph{Automated Software Engineering}, vol.~29, no.~1, pp. 1--32, 2022.

\bibitem{garcia2020comprehensive}
\BIBentryALTinterwordspacing
G.~Joshua, F.~Yang, S.~Junjie, A.~Sumaya, X.~Yuan, Chen, and Q.~Alfred, ``A
  comprehensive study of autonomous vehicle bugs,'' in \emph{Proceedings of the
  ACM/IEEE 42nd International Conference on Software Engineering}, ser. ICSE
  '20.\hskip 1em plus 0.5em minus 0.4em\relax New York, NY, USA: Association
  for Computing Machinery, 2020, p. 385–396. [Online]. Available:
  \url{https://doi.org/10.1145/3377811.3380397}
\BIBentrySTDinterwordspacing

\bibitem{zhang2019}
T.~Zhang, C.~Gao, L.~Ma, M.~Lyu, and M.~Kim, ``An empirical study of common
  challenges in developing deep learning applications,'' in \emph{2019 IEEE
  30th International Symposium on Software Reliability Engineering
  (ISSRE)}.\hskip 1em plus 0.5em minus 0.4em\relax 3 Park Avenue, New York NY
  10016-5997, USA: IEEE, 2019, pp. 104--115.

\bibitem{wirsansky2020hands}
E.~Wirsansky, \emph{Hands-on genetic algorithms with Python: applying genetic
  algorithms to solve real-world deep learning and artificial intelligence
  problems}.\hskip 1em plus 0.5em minus 0.4em\relax Packt Publishing Ltd: Packt
  Publishing Ltd, 2020.

\bibitem{github_issue_bot_generate}
RasaHQ, \url{https://github.com/RasaHQ/rasa/issues/8541}, 2021, accessed:
  2021-11-01.

\bibitem{github_issue_bot_close}
------, \url{https://github.com/RasaHQ/rasa/issues/4730}, 2021, accessed:
  2021-11-01.

\bibitem{issue_sample_08}
``some tests fails with pytorch 1.6,''
  \url{https://github.com/speechbrain/speechbrain/issues/248}, 2020.

\bibitem{issue_sample_09}
``Add typing for trainer.logger,''
  \url{https://github.com/Lightning-AI/lightning/pull/11114}, 2021.

\bibitem{issue_sample_10}
``Loading a checkpoint that was saved in pl < 1.2 still breaks,''
  \url{https://github.com/Lightning-AI/lightning/issues/7400}, 2021.

\bibitem{issue_sample_11}
``reduce\_lr\_on\_plateau can't find validation metrics in most recent
  release.'' \url{https://github.com/scverse/scvi-tools/issues/1112}, 2021.

\bibitem{issue_sample_12}
``Bug: Mismatch value with speechbrain.nnet.pooling.statisticalpooling,''
  \url{https://github.com/speechbrain/speechbrain/issues/1048}, 2021.

\bibitem{issue_sample_13}
``Bug fix for static pulse shapes,''
  \url{https://github.com/cms-sw/cmssw/pull/23001}, 2018.

\bibitem{issue_sample_14}
``Fix docs for early stopping,''
  \url{https://github.com/Lightning-AI/lightning/pull/865}, 2020.

\bibitem{10.1145/1117696.1117704}
\BIBentryALTinterwordspacing
J.~Anvik, L.~Hiew, and G.~C. Murphy, ``Coping with an open bug repository,'' in
  \emph{Proceedings of the 2005 OOPSLA Workshop on Eclipse Technology
  EXchange}, ser. eclipse '05.\hskip 1em plus 0.5em minus 0.4em\relax New York,
  NY, USA: Association for Computing Machinery, 2005, p. 35–39. [Online].
  Available: \url{https://doi.org/10.1145/1117696.1117704}
\BIBentrySTDinterwordspacing

\bibitem{Long22}
\BIBentryALTinterwordspacing
G.~Long and T.~Chen, ``On reporting performance and accuracy bugs for deep
  learning frameworks: An exploratory study from github,'' 2022. [Online].
  Available: \url{https://arxiv.org/abs/2204.07893}
\BIBentrySTDinterwordspacing

\bibitem{menzies2019five}
T.~Menzies, ``The five laws of se for ai,'' \emph{IEEE Software}, vol.~37,
  no.~1, pp. 81--85, 2019.

\bibitem{jia2021symptoms}
L.~Jia, H.~Zhong, X.~Wang, L.~Huang, and X.~Lu, ``The symptoms, causes, and
  repairs of bugs inside a deep learning library,'' \emph{Journal of Systems
  and Software}, vol. 177, p. 110935, 2021.

\bibitem{issue_sample_04}
``Loading saved destvi.from\_rna\_model dosn't consider two anndatas used to
  create model,'' \url{https://github.com/scverse/scvi\-tools/issues/1087},
  2021.

\bibitem{issue_sample_05}
``Fix model architecture for deployment to onnx,''
  \url{https://github.com/mehta-lab/microDL/pull/234}, 2023.

\bibitem{issue_sample_06}
``tf.function not used for model inference,''
  \url{https://github.com/RasaHQ/rasa/issues/10728}, 2022.

\bibitem{issue_sample_07}
``Fix default ckpt path when logger exists,''
  \url{https://github.com/Lightning\-AI/lightning/pull/771}, 2020.

\bibitem{kononenko2018studying}
O.~Kononenko, T.~Rose, O.~Baysal, M.~Godfrey, D.~Theisen, and B.~De~Water,
  ``Studying pull request merges: a case study of shopify's active merchant,''
  in \emph{Proceedings of the 40th International Conference on Software
  Engineering: Software Engineering in Practice}.\hskip 1em plus 0.5em minus
  0.4em\relax 3 Park Avenue, New York NY 10016-5997, USA: IEEE, 2018, pp.
  124--133.

\bibitem{chaturvedi2014predicting}
K.~Chaturvedi, P.~Kapur, S.~Anand, and V.~Singh, ``Predicting the complexity of
  code changes using entropy based measures,'' \emph{International Journal of
  System Assurance Engineering and Management}, vol.~5, no.~2, pp. 155--164,
  2014.

\bibitem{arcuri2014hitchhiker}
A.~Arcuri and L.~Briand, ``A hitchhiker's guide to statistical tests for
  assessing randomized algorithms in software engineering,'' \emph{Software
  Testing, Verification and Reliability}, vol.~24, no.~3, pp. 219--250, 2014.

\bibitem{Kampenes07}
\BIBentryALTinterwordspacing
V.~B. Kampenes, T.~Dybå, J.~E. Hannay, and D.~I. Sjøberg, ``A systematic
  review of effect size in software engineering experiments,''
  \emph{Information and Software Technology}, vol.~49, no.~11, pp. 1073--1086,
  2007. [Online]. Available:
  \url{https://www.sciencedirect.com/science/article/pii/S0950584907000195}
\BIBentrySTDinterwordspacing

\bibitem{Cliff93}
N.~Cliff, ``Dominance statistics: Ordinal analyses to answer ordinal
  questions.'' \emph{Psychological bulletin}, vol. 114, no.~3, p. 494, 1993.

\bibitem{macbeth2011cliff}
G.~Macbeth, E.~Razumiejczyk, and R.~D. Ledesma, ``Cliff's delta calculator: A
  non-parametric effect size program for two groups of observations,''
  \emph{Universitas Psychologica}, vol.~10, no.~2, pp. 545--555, 2011.

\bibitem{Romano06}
J.~Romano, J.~D. Kromrey, J.~Coraggio, J.~Skowronek, and L.~Devine, ``Exploring
  methods for evaluating group differences on the nsse and other surveys: Are
  the t-test and cohen’sd indices the most appropriate choices,'' in
  \emph{annual meeting of the Southern Association for Institutional
  Research}.\hskip 1em plus 0.5em minus 0.4em\relax The Pennsylvania State
  University: Citeseer, 2006, pp. 1--51.

\bibitem{lda-gensim}
``Gensim library,'' \url{https://radimrehurek.com/gensim_3.8.3/index.html},
  2022.

\bibitem{mallet}
``Mallet library,'' \url{https://mimno.github.io/Mallet/}, 2022.

\bibitem{romano2021empirical}
A.~Romano, X.~Liu, Y.~Kwon, and W.~Wang, ``An empirical study of bugs in
  webassembly compilers,'' in \emph{2021 36th IEEE/ACM International Conference
  on Automated Software Engineering (ASE)}.\hskip 1em plus 0.5em minus
  0.4em\relax 3 Park Avenue, New York NY 10016-5997, USA: IEEE, 2021, pp.
  42--54.

\bibitem{bosu2014impact}
A.~Bosu and J.~C. Carver, ``Impact of developer reputation on code review
  outcomes in oss projects: An empirical investigation,'' in \emph{Proceedings
  of the 8th ACM/IEEE international symposium on empirical software engineering
  and measurement}, 2014, pp. 1--10.

\bibitem{schober2018correlation}
P.~Schober, C.~Boer, and L.~A. Schwarte, ``Correlation coefficients:
  appropriate use and interpretation,'' \emph{Anesthesia \& Analgesia}, vol.
  126, no.~5, pp. 1763--1768, 2018.

\bibitem{tagra2022revisiting}
A.~Tagra, H.~Zhang, G.~K. Rajbahadur, and A.~E. Hassan, ``Revisiting reopened
  bugs in open source software systems,'' \emph{Empirical Software
  Engineering}, vol.~27, no.~4, pp. 1--34, 2022.

\bibitem{zimmermann2012characterizing}
T.~Zimmermann, N.~Nagappan, P.~J. Guo, and B.~Murphy, ``Characterizing and
  predicting which bugs get reopened,'' in \emph{2012 34th International
  Conference on Software Engineering (ICSE)}.\hskip 1em plus 0.5em minus
  0.4em\relax IEEE, 2012, pp. 1074--1083.

\bibitem{bosu2016process}
A.~Bosu, J.~C. Carver, C.~Bird, J.~Orbeck, and C.~Chockley, ``Process aspects
  and social dynamics of contemporary code review: Insights from open source
  development and industrial practice at microsoft,'' \emph{IEEE Transactions
  on Software Engineering}, vol.~43, no.~1, pp. 56--75, 2016.

\bibitem{maddila2019predicting}
C.~Maddila, C.~Bansal, and N.~Nagappan, ``Predicting pull request completion
  time: a case study on large scale cloud services,'' in \emph{Proceedings of
  the 2019 27th acm joint meeting on european software engineering conference
  and symposium on the foundations of software engineering}, 2019, pp.
  874--882.

\bibitem{tan2014bug}
L.~Tan, C.~Liu, Z.~Li, X.~Wang, Y.~Zhou, and C.~Zhai, ``Bug characteristics in
  open source software,'' \emph{Empirical software engineering}, vol.~19, pp.
  1665--1705, 2014.

\bibitem{nikanjam2021automatic}
A.~Nikanjam, H.~B. Braiek, M.~M. Morovati, and F.~Khomh, ``Automatic fault
  detection for deep learning programs using graph transformations,'' \emph{ACM
  Transactions on Software Engineering and Methodology (TOSEM)}, vol.~31,
  no.~1, pp. 1--27, 2021.

\bibitem{zhang2021autotrainer}
X.~Zhang, J.~Zhai, S.~Ma, and C.~Shen, ``Autotrainer: An automatic dnn training
  problem detection and repair system,'' in \emph{2021 IEEE/ACM 43rd
  International Conference on Software Engineering (ICSE)}.\hskip 1em plus
  0.5em minus 0.4em\relax IEEE, 2021, pp. 359--371.

\bibitem{wardat2022deepdiagnosis}
M.~Wardat, B.~D. Cruz, W.~Le, and H.~Rajan, ``Deepdiagnosis: automatically
  diagnosing faults and recommending actionable fixes in deep learning
  programs,'' in \emph{Proceedings of the 44th international conference on
  software engineering}, 2022, pp. 561--572.

\bibitem{tambon2021silent}
F.~Tambon, A.~Nikanjam, L.~An, F.~Khomh, and G.~Antoniol, ``Silent bugs in deep
  learning frameworks: An empirical study of keras and tensorflow,'' 2021.

\bibitem{hu2021model}
X.~Hu, L.~Chu, J.~Pei, W.~Liu, and J.~Bian, ``Model complexity of deep
  learning: A survey,'' \emph{Knowledge and Information Systems}, vol.~63, pp.
  2585--2619, 2021.

\bibitem{yao2017complexity}
Y.~Yao, Z.~Xiao, B.~Wang, B.~Viswanath, H.~Zheng, and B.~Y. Zhao, ``Complexity
  vs. performance: empirical analysis of machine learning as a service,'' in
  \emph{Proceedings of the 2017 Internet Measurement Conference}, 2017, pp.
  384--397.

\bibitem{bennett2000software}
K.~H. Bennett and V.~T. Rajlich, ``Software maintenance and evolution: a
  roadmap,'' in \emph{Proceedings of the Conference on the Future of Software
  Engineering}, 2000, pp. 73--87.

\bibitem{cao2021characterizing}
\BIBentryALTinterwordspacing
J.~Cao, B.~Chen, C.~Sun, L.~Hu, and X.~Peng, ``Characterizing performance bugs
  in deep learning systems,'' 2021. [Online]. Available:
  \url{https://arxiv.org/abs/2112.01771}
\BIBentrySTDinterwordspacing

\bibitem{liu2021detecting}
C.~Liu, J.~Lu, G.~Li, T.~Yuan, L.~Li, F.~Tan, J.~Yang, L.~You, and J.~Xue,
  ``Detecting tensorflow program bugs in real-world industrial environment,''
  in \emph{2021 36th IEEE/ACM International Conference on Automated Software
  Engineering (ASE)}.\hskip 1em plus 0.5em minus 0.4em\relax 3 Park Avenue, New
  York NY 10016-5997, USA: IEEE, 2021, pp. 55--66.

\bibitem{rivera2021challenge}
E.~Rivera-Landos, F.~Khomh, and A.~Nikanjam, ``The challenge of reproducible
  ml: an empirical study on the impact of bugs,'' 2021.

\bibitem{chen2015mxnet}
\BIBentryALTinterwordspacing
T.~Chen, M.~Li, Y.~Li, M.~Lin, N.~Wang, M.~Wang, T.~Xiao, B.~Xu, C.~Zhang, and
  Z.~Zhang, ``Mxnet: A flexible and efficient machine learning library for
  heterogeneous distributed systems,'' 2015. [Online]. Available:
  \url{https://arxiv.org/abs/1512.01274}
\BIBentrySTDinterwordspacing

\bibitem{jia2014caffe}
\BIBentryALTinterwordspacing
Y.~Jia, E.~Shelhamer, J.~Donahue, S.~Karayev, J.~Long, R.~Girshick,
  S.~Guadarrama, and T.~Darrell, ``Caffe: Convolutional architecture for fast
  feature embedding,'' in \emph{Proceedings of the 22nd ACM International
  Conference on Multimedia}, ser. MM '14.\hskip 1em plus 0.5em minus
  0.4em\relax New York, NY, USA: Association for Computing Machinery, 2014, p.
  675–678. [Online]. Available: \url{https://doi.org/10.1145/2647868.2654889}
\BIBentrySTDinterwordspacing

\bibitem{hanam2016discovering}
Q.~Hanam, F.~S. d.~M. Brito, and A.~Mesbah, ``Discovering bug patterns in
  javascript,'' in \emph{Proceedings of the 2016 24th ACM SIGSOFT international
  symposium on foundations of software engineering}, 2016, pp. 144--156.

\end{thebibliography}

\end{document}